\documentstyle[12pt,aas2pp4]{article}

\newcommand{\kms}{$\,\mbox{km}\,\mbox{s}^{-1}$}
\newcommand{\ea}{\emph{et al.\ }}

\slugcomment{submitted to AJ}

\lefthead{Walter \& Brinks}
\righthead{Holes and Shells in IC~2574}

\begin{document}

\title{Holes and Shells in the Interstellar Medium of the Nearby Dwarf
Galaxy IC 2574} 
 
\author{Fabian Walter\footnote{Visiting Astronomer, German--Spanish
Astronomical Centre, Calar Alto, operated by the
Max--Planck--Institute for Astronomy, Heidelberg, jointly with the
Spanish Commission for Astronomy.}}  \affil{Radioastronomisches
Institut der Universit\"{a}t, Auf dem H\"{u}gel 71, D--53121 Bonn,
Germany
\\Electronic mail: walter@astro.uni-bonn.de} 

\and

\author{Elias Brinks} \affil{Departamento de Astronom\'{\i}a, Apartado
Postal 144, Guanajuato, Gto. 36000, Mexico\\National Radio Astronomy
Observatory, P.O. Box O, Socorro, NM 87801 \\Electronic mail:
ebrinks@astro.ugto.mx}

\begin{abstract}

We present \ion{H}{1} synthesis observation of the nearby dwarf galaxy
IC~2574 (a member of the M81 group of galaxies) made with the NRAO
Very Large Array in its B--, C-- and D--array configurations at high
spatial and velocity resolution (95 pc $\times$ 2.6 \kms). In
addition, we present optical broad and narrow band images obtained
with the 2.2--m telescope of the Calar Alto observatory. The VLA
\ion{H}{1} observations show a stunning amount of detail in the form
of 48 mostly expanding \ion{H}{1} shells and holes in its neutral
interstellar medium. These features range in size from about 100 (a
limit set by the size of the beam) to about 1000 pc dominating the
appearance of the \ion{H}{1} surface brightness map. Their dynamics
clearly influence the velocity field of IC~2574. In addition to well
defined holes, some large scale ($> 1000$ pc) coherent features are
visible in the channel maps. They are probably the remainder of an
older shell population.

Current star formation, as traced by H$\alpha$ emission, is
predominantly found along the rims of the larger \ion{H}{1} holes,
suggesting propagating star formation. On linear scales of $\approx
95$~pc, star formation occurs if the \ion{H}{1} surface density
reaches values higher than $10^{21}$~cm$^{-2}$. The radial expansion
of the \ion{H}{1} holes (8--12 \kms), the indicative ages (10--60
$10^6$ yr) and the energy requirements for their formation
($10^{50}$--$10^{53}$ ergs) can be understood in terms of the combined
effects of stellar winds and multiple supernova explosions of the most
massive stars formed during a recent phase of active star formation.

The scaleheight of the \ion{H}{1} layer is found to be $\approx
350$~pc, considerably thicker than that in massive disk galaxies. This
is due to a lower gravitational potential (for the same observed
onedimensional velocity dispersion of $\approx 7$ \kms). This puffed
up disk implies a lower \ion{H}{1} volume density (0.15
cm$^{-3}$). This, combined with the reduced gravitational pull and
solid body rotation throughout the disk explains why the diameter
distribution in dwarf galaxies extends to substantially larger values
than in spiral galaxies.

A comparison with other galaxies shows that the energies needed to
create these structures are the same for all types of galaxies, at
least to first order.  The overall statistical properties of the
\ion{H}{1} holes and shells in galaxies show clear trends with Hubble
type (or rather mass), such as in their diameter distribution,
expansion velocities and ages.

\end{abstract}



\section{Introduction}
\label{Intro}

Some twenty years ago, Heiles (\markcite{79}1979, \markcite{84}1984,
\markcite{90}1990) discovered large, shell--like structures in the
neutral hydrogen (\ion{H}{1}) distribution of our Galaxy by inspection
of maps made in the 21--cm line. Since this discovery, a wealth of
observations obtained with powerful Synthesis Radio Telescopes such as
the Very Large Array (VLA) or the Westerbork Synthesis Radio Telescope
(WSRT) revealed similar structures in our nearest neighbors. Prominent
examples are M~31 (Brinks \markcite{BRI81}1981, Brinks \& Bajaja
\markcite{BRI86}1986), M~33 (Deul \& den Hartog \markcite{DEU90}1990),
Holmberg~II (Puche \ea \markcite{PUC92}1992, hereafter referred to as
PWBR92), the galaxies M~101 and NGC~6946 (Kamphuis
\markcite{KAM93}1993) and IC~10 (Wilcots \& Miller 
\markcite{WIL89}1998). The latest, impressive additions to this list
are the SMC (Staveley--Smith \ea \markcite{STA97}1997) and LMC (Kim
\ea \markcite{KIM98}1998). All these observations indicate that the
interstellar medium (ISM) of medium to late--type galaxies is
dominated by features, which are variously described as shells, rings,
holes, loops, bubbles or cavities.

Obviously, one of the basic questions is which physical process is
able to produce these observed structures.  The general picture which
has emerged is that they are the result of combined stellar winds and
supernova explosions produced by young stellar associations. For
review articles on that topic see Tenorio--Tagle \& Bodenheimer
\markcite{TEN88}(1988), and van der Hulst \markcite{HUL96}(1996, and
references therein). Based on a simple model of holes created by O and
B stars, Oey \& Clarke \markcite{OEY97}(1997) successfully predict the
observed number distribution of SMC holes, lending support to this
hypothesis. It should be noted, however, that this general picture is
not without its flaws. Searches for the remnant stellar populations of
the OB associations thought to be responsible for some of the holes in
Ho~II have not led to the expected result (Rhode \ea
\markcite{RHO97}1999). And in the case of the largest observed shells,
these seem to have energy requirements surpassing the output of
stellar winds and supernovae. To explain those structures an
alternative mechanism was proposed, the infall of gas clouds (see
Tenorio--Tagle \ea \markcite{TEN87}(1987) for a numerical simulation
and van der Hulst \& Sancisi \markcite{HUL88}(1988) for observational
evidence in the case of one of the largest holes in M~101). And
recently, even Gamma Ray Bursters (GRBs) have been suggested as a
possible explanation for the largest \ion{H}{1} holes (Efremov,
Elmegreen \& Hodge \markcite{EFR98}1998, Loeb \& Perna
\markcite{LOE98}1998).

The idea that supernovae deposit large amounts of energy into the ISM
is not new. Cox \& Smith \markcite{COX74}(1974) already discussed the
idea that the energy released by supernovae sets up tunnels of hot gas
in the ISM. This picture was a substantial modification of the Field,
Goldsmith \& Habing \markcite{FIE69}(1969) model which consists of a
two phase medium. The ideas proposed by Cox \& Smith were incorporated
within the model elaborated by McKee \& Ostriker \markcite{MCK77}(1977)
who suggested that the ISM has three distinct phases, a cool
(T$\,\approx10^2\,$~K), a warm (T$\,\approx10^4\,$~K), and a hot phase
(T$\,\approx10^6\,$~K), all in pressure equilibrium. Various
refinements have been proposed but this picture basically still stands.

Knowing how much kinetic energy is transferred in expanding shells and
filaments is of direct importance for assessing the energy balance in
the ISM. In addition to an analytical approach, various authors are
pushing forward numerical simulations (e.g., Palou\v{s}, Franco \&
Tenorio--Tagle \markcite{PAL90}1990, Silich \ea \markcite{SIL96}1996,
Palou\v{s}, Tenorio--Tagle \& Franco \markcite{PAL94}1994), in order
to model the evolution of the shape of the \ion{H}{1} shells in
various environments, the fragmentation of matter on the rim of the
shells and to investigate induced star formation. This is important
not only for tracking down the mechanisms which produce the
\ion{H}{1}\ shells and filaments, but also for determining what
fraction of the ISM breaks out into the halo and might eventually rain
back onto the disk.

Our aim is to investigate the properties of the \ion{H}{1}\ shells and
bubbles in the ISM of galaxies. Intimately linked to this are the
questions when fragmentation on the rim of the shells starts and what
drives star formation. Unfortunately, the determination of the
characteristics of \ion{H}{1} shells in our Galaxy is not
straightforward due to our unfavorable location within the disk.
Looking at the nearest spiral galaxies (e.g., M~31, M~33, M~101,
NGC~6946) does not improve matters all that much since the structures
in their ISM are strongly modified by spiral density waves and
differential rotation, both effects dramatically influencing the shape
and formation of holes and shells. Therefore, we decided to direct our
attention to nearby dwarf galaxies instead. Dwarfs are slow rotators,
generally show solid body rotation, and lack density waves. This
implies that once features like shells have formed, they won't be
deformed by galactic shear and therefore tend to be long
lived. Moreover, the overall gravitational potential of a dwarf is
much smaller than in a normal spiral. The same amount of energy input
of a star forming region therefore has a more pronounced impact on the
overall appearance of the ISM, as shown by PWBR92. Since dwarf
galaxies also tend to have a much thicker disk compared to massive
spirals (e.g., Staveley--Smith, Davies \& Kinman
\markcite{STA92}1992), the \ion{H}{1}\ volume density of dwarf
galaxies is lower, again facilitating the creation of large holes. And
because of their smaller size, the probability to suffer impacts from
infalling clouds is lower compared to that for larger
galaxies. Despite all these arguments in favor, only very few such
\ion{H}{1} studies of dwarf galaxies have been published. Beside our
two neighbors, the SMC and LMC a detailed analysis was performed in
one case only, that of the dwarf galaxy Holmberg~II (PWBR92), member
of the M~81 group.

In this paper, we concentrate on IC~2574, a gas--rich dwarf galaxy in
the same group. Table~\ref{ic_info} summarizes some general
information on this object. IC~2574 (= UGC~5666 = DDO~81, also
sometimes referred to as Coddington's Nebula) has been studied by
several authors in the past. Observations with single dish telescopes
in the 21--cm line of neutral hydrogen (\ion{H}{1}) were published by
Rots \markcite{ROT80}(1980) and Huchtmeier \& Richter
\markcite{HUC88}(1988) and a first study using an interferometer was
made by Seielstad \& Wright \markcite{SEI73}(1973). A recent
\ion{H}{1} study, concentrating on the dynamics and contribution to
the total mass by dark matter was performed by Martimbeau, Carignan \&
Roy \markcite{MAR94}(1994, hereafter abbreviated MCR94), who observed
the galaxy with the Westerbork Synthesis Radio Telescope.  Recent
broadband optical studies were published by Tikhonov \ea
\markcite{TIK91}(1991). Studies of \ion{H}{2} regions in IC~2574 have
been presented by Miller \& Hodge (\markcite{MIL94}1994), Kennicutt
\markcite{KEN88}(1988) and Arsenault \& Roy
\markcite{ARS86}(1986). Spectroscopy of some prominent \ion{H}{2}
regions was performed by Miller \& Hodge (\markcite{MIL96}1996) and
H$\alpha$ velocity fields were recently obtained by Tomita \ea\
(\markcite{TOM98}1998). In addition, four Wolf--Rayet stars were
discovered by Drissen, Roy \& Moffat (\markcite{DRI93}1993).

\placetable{ic_info}

We present \ion{H}{1}\ observations made with the
NRAO\footnote{The National Radio Astronomy Observatory (NRAO) is
operated by Associated Universities, Inc., under cooperative agreement
with the National Science Foundation.} Very Large Array (VLA) in its
B--, C-- \& D--array configurations supplemented by optical
observations obtained at the Calar Alto Observatory. 
The VLA data were originally obtained by Puche, Brinks
and Westpfahl with the intention of following up on their work on Ho
II. So far, only a low resolution integrated \ion{H}{1} map was
published by Puche \& Westpfahl \markcite{PUC94}(1994). This paper
presents the VLA data in a complete form and compares them with
optical observations.  The original data were recalibrated by us to
produce far superior maps using the latest calibration and imaging
techniques within {\sc aips}. The data reduction and results are discussed
in Secs.~\ref{obser}~\&~\ref{histu}.  In Sec.~\ref{hihol}, a catalog
of \ion{H}{1} holes and shells is presented. Sec.~\ref{ha-hi}
discusses the relation between \ion{H}{1} and H$\alpha$ features.  A
general discussion is given in Sec.~\ref{discu} and a summary of the
results is presented in Sec.~\ref{summa}.

\section{Observations}
\label{obser}

\subsection{Radio Observations}

IC~2574 was observed with the VLA in B--, C-- and
D--configuration. Part of the D--array observations were affected by
solar interference, which was removed by deleting {\it uv--}spacings
shorter than typically 0.4 k$\lambda$ and editing out bad
time--ranges.  In total, 17 hours were spent on source which were
divided into 4 hours for D--array, 2.5 hours in C--array and 10.5
hours in B--array (see Table~\ref{VLAsetup} for a detailed description
of the VLA observations). The longer observing time in B--array only
partly compensates for the lower surface brightness sensitivity in
this extended configuration. The absolute flux calibration was
determined by observing 1328+307 (3C286) for approximately 20 minutes
during each observing run, assuming a flux density of 14.73 Jy
according to the Baars \ea \markcite{BAA77}(1977) scale. This
calibrator was also used to derive the complex bandpass
corrections. The nearby calibrators 1031+567 and 0945+664 were used as
secondary amplitude and phase calibrators and their fluxes were
determined to be 1.80 Jy and 2.22 Jy, respectively. Since the systemic
velocity of IC~2574 ($\approx 53$ \kms) is close in velocity to
\ion{H}{1} emission from our Galaxy, each calibrator observation was
split in two parts to form a pair of observations, one part having the
velocity shifted by +300 \kms\ and the other by -300 \kms\ to avoid
contamination by Galactic emission. In the course of the calibration
these observations were then averaged to give interpolated amplitude
and phase corrections. IC~2574 was observed using a 1.56 MHz bandwidth
centered at a heliocentric velocity of 38 \kms. This band was divided
into 128 channels resulting in a velocity resolution of 2.58 \kms\
after online Hanning smoothing.

The data for each array were edited and calibrated separately with the
{\sc aips} package\footnote{The Astronomical Image Processing System
({\sc aips}) has been developed by the NRAO.}. The {\it uv--}data were
inspected and bad data points due to either interference or cross talk
between antennae were removed, after which the data were
calibrated. We Fourier transformed our B--, C-- and D--array
observations separately to assess their quality; subsequently we
combined all data to form a single dataset which was used for
mapping. All the results presented in this paper are based on this
combined dataset.

Channels with velocities lower then $-40$ \kms\ and higher than $+136$
\kms\ were found to be line free; these channels determined the
continuum emission which was subtracted in the {\it uv--}plane. The same
channels were used to produce a radio continuum map of IC~2574.

After that, two sets of datacubes (1024 $\times$ 1024 pixels $\times$
68 channels each) were produced using the task {\sc imagr} in {\sc
aips}, each of them {\sc clean}ed to a level of two times the rms
noise (H\"ogbom \markcite{HOG74}1974, Clark \markcite{CLA80}1980). The
first data set was made with natural weighting, leading to a
resolution of $12.4'' \times 11.9''$ and emphasizing large scale
structures. A second cube was produced using the {\sc robust}
weighting scheme (Briggs \markcite{BRI95}1995). This scheme achieves
the high sensitivity of natural weighting combined with a well behaved
synthesized beam, at a resolution close to that of uniform
weighting. To obtain the optimum {\sc robust} value for our
observation, the {\sc robust} parameter space was searched by
producing maps in the sensitive regime of $-2<${\sc robust}$<2$ and
measuring for each map the rms noise and beamsize. Eventually, we
chose a value of {\sc robust}=0.25, resulting in a beamsize of $6.4''
\times 5.9''$ and an rms noise of 0.7 mJy beam$^{-1}$. A comparison
with the beamsize of the high resolution uniform map ($4.7''\times
4.3''$) and the rms noise of the low--noise, natural map ($\approx 10
\%$ increase in noise) shows the strength of this mapping
algorithm. Unless otherwise mentioned, the results presented in the
following are based on this {\sc robust} cube.

Although IC~2574 is situated at a galactic latitude of 43.6$^{\circ}$,
some of the channels were clearly confused by Galactic emission (in
particular between $+10$ and $-10$ \kms). Since galactic emission
shows its presence on predominantly large scales, we were able to
remove the most prominent emission features by blanking the {\it uv--}data
out to 0.1 k$\lambda$ in these channels.

To separate real emission from the noise, the following procedure was
applied: the natural weighted data cube was convolved to a circular
beam with a FWHM of $45''$. The smoothed map was then tested at the
$2\sigma$ level; if a pixel fell below this level, the counterpart in
the cube was blanked. After that, the remaining peaks were
inspected. Emission that was present in 3 consecutive channels was
considered to be real while all other remaining spikes were considered
to be noise and blanked. The final result was named the master
cube. This master cube can be used as a mask to blank the natural and
{\sc robust} data cubes. This method ensures that the same regions are
included when inspecting cubes at different resolutions and with
different signal--to--noise ratios.

\placetable{VLAsetup}

\subsection{Optical Observations}
\label{optical}

IC~2574 was observed with the 2.2--m telescope of the Calar Alto
Observatory\footnote{The Calar Alto Observatory is operated by the
Max--Planck--Institute for Astronomy, Heidelberg, jointly with the
Spanish Commission for Astronomy.} on 1999 January 25 and 26 in
Johnson R--band (20 min) as well as in H$\alpha$ (45 min).  A $2048
\times 2048$ pixel CCD was used leading to $0.5''$ per pixel. The
seeing was around $1.3''$. A focal reducer (CAFOS) allowed for a field
of view of radius $8'$ to be imaged. The usual calibration steps were
followed using the {\sc iraf} package\footnote{{\sc iraf} is
distributed by National Optical Astronomy Observatories, which is
operated by the Association of Universities for Research in Astronomy
(AURA), Inc., under contract with the National Science Foundation.}.
Each exposure was corrected for zero offset (bias) and was flatfielded
using skyflats. After that, frames were inspected for bad pixels and
shifted to get them aligned.  Lastly, the exposures were combined to
eliminate cosmic ray hits and to create a final image.  As it turned
out, conditions were not photometric during the observation, hence no
calibration using standard stars was performed. In order to get an
approximate calibration for IC~2574, values were taken from the
literature (\markcite{MAR94}MCR94). The deep R--band image was used
for the continuum subtraction of the H$\alpha$ map.  The continuum
level was determined by simply scaling the continuum image relative to
the line plus continuum image such that the foreground stars
disappeared when the two images were subtracted. For the calibration
we relied on the H$\alpha$ fluxes published by Miller \& Hodge
\markcite{MIL94}(1994, hereafter referred to as MH94).  In their
paper, these authors identify 289 \ion{H}{2} regions in IC~2574 (for a
thorough discussion on the H$\alpha$ emission and its calibration see
the Appendix in \markcite{MIL94}MH94). Figs.~\ref{ic_r}
and~\ref{ic_ha} show our R--band and the H$\alpha$ image,
respectively.

Coordinates for the maps were obtained using the task {\sc koords}
within the {\sc karma} software package (for a description of {\sc
karma} see Sec.~\ref{hicat}). We used as input an image from the
Digital Sky Survey. Using 14 stars found in our CCD image, it was
possible to determine the coordinate system to an accuracy of better
than one arcsecond, more than adequate for our needs.

\placefigure{ic_r}

\placefigure{ic_ha}

\section{HI Studies}
\label{histu}

\subsection{Global \ion{H}{1}\ Properties}

The channel maps obtained by the robust weighted cube are presented in
Fig.~\ref{ic_cm_1}.  As can be seen, most traces of contamination by
Galactic emission were successfully removed. To save space, only every
other channel over the frequency range where line emission is present
is shown. The heliocentric radial velocities (in \kms) are plotted in
the upper right--hand corner of each plane. The beamsize ($6.4'' \times
5.9''$) is indicated in the top left--hand panel (note that the
beamsize is so small that it cannot be properly reproduced in our
maps). Many holes and shell--like structures are visible in the channel
maps; these features will be the main topic of this paper.

Obtaining the \ion{H}{1} flux of each channel and hence the global
\ion{H}{1}\ profile of IC~2574 is not straightforward. As first
pointed out by J\"ors\"ater \& van Moorsel \markcite{JOR95}(1995),
fully cleaned maps do not exist and any cleaned map consists of the
sum of two maps: one containing the restored clean components and the
other the residual map. In the former, the unit is Jansky per clean
beam area, and in the latter Jansky per dirty beam area. Usually,
fluxes are determined on a cleaned map, assuming that the clean beam
is the correct one for the entire map. In reality, the flux is
calculated correctly only for the cleaned fraction of the map; for the
residual map the flux is overestimated when the dirty beam area is
bigger than the clean beam area (which is usually the case).  This
becomes a real problem for extended objects, especially in the case of
combined array data, like for our VLA data set. For a full discussion
on this topic see J\"ors\"ater \& van Moorsel \markcite{JOR95}(1995).
Following their prescription, the real flux of a channel map is given
by:
$$
G=\frac{D \times C}{D-R}
$$
where $C$ is the cleaned flux, and $R$ and $D$ are the 'erroneous'
residual flux and the 'erroneous' dirty flux over the same area of a
channel map, respectively. We followed this procedure for every single
channel to obtain the global \ion{H}{1}\ profile as presented in
Fig.~\ref{ic_glob}. Without this correction, we would have
overestimated the \ion{H}{1} flux by a factor of about two!

Figure~\ref{ic_glob} shows the flux-- and primary--beam corrected
fluxes for every channel of IC~2574 (open squares) plotted together
with the single dish profile published by Rots \markcite{ROT80}(1980)
and the WSRT data by MCR94\markcite{MAR94} (the arrows indicate the
velocity range for which the data points were interpolated due to
contamination by galactic emission). From our data, we derive a total
integrated \ion{H}{1}\ flux of 373 Jy \kms\ which should be compared
with the value of 399 Jy \kms\ obtained by Rots
\markcite{ROT80}(1980), who used the Greenbank 300 ft telescope to map
IC~2574. This is excellent agreement, given the uncertainties of
removing Galactic emission and systematic errors in the absolute
calibration. This also means that we hardly miss any \ion{H}{1} flux
due to missing short spacings. Other values obtained by
interferometric observations are: $440\pm 50$ Jy \kms\ found by
Seielstad \& Wright \markcite{SEI73}(1973), using the Owens Valley
Radio Interferometer and 286 Jy \kms\ obtained by MCR94 from their
Westerbork observations. As can be seen from Fig.~\ref{ic_glob}, the
new VLA observations recover substantially more flux than the WSRT
data. Adopting a distance of 3.2 Mpc, our flux translates to a total
neutral hydrogen mass of $M_{HI}=(9.0\pm 0.1)\times 10^8\,
\mbox{M}_{\odot}$ for IC~2574.

\placefigure{ic_cm_1}

\notetoeditor{please note that Figure 3 -- the channel maps --
actually consists of 3 Figures (a, b and c) -- the filenames are
walter.fig3an.ps, walter.fig3bn.ps and walter.fig3cn.ps)}

\placefigure{ic_glob}

\subsection{Global Characteristics}
\label{global}

Fig.~\ref{ic_mom0} shows the \ion{H}{1} surface density map of IC~2574
obtained by adding all channels containing \ion{H}{1} emission of the
robust cube, after blanking them with the master cube. Note that the
beamsize is so small that it can hardly be properly reproduced. Note
also that, although many hole--like structures and cavities are
visible, specific features turn out to be less prominent in the total
surface brightness map than in the channel maps. This is related to
the fact that the galaxy is highly inclined, leading to overlapping of
shell--like structures along the line--of--sight in the integrated
map. For a more complete discussion on that topic, see
Sec.~\ref{hicat}.

Fig.~\ref{ic_mom1} shows the velocity field of IC~2574 which was
calculated in the usual manner by taking the first moment. Since the
high resolution velocity field is dominated by small--scale motions
within the ISM, the natural weighted cube was convolved to $30''$ (as
indicated by the beamsize) to put more emphasis on the global
characteristics of the velocity field. The underlying greyscale is a
linear representation of the \ion{H}{1} surface brightness map at full
resolution. Although the overall velocity field shows the typical
behavior for a dwarf galaxy (solid body rotation in the center,
flattening of the rotation curve towards the very edge of the galaxy),
the general rotation is obviously disturbed by noncircular motions
(e.g., note the prominent band of lower velocities to the north east
of the center of the galaxy, parallel to its major axis). These
deviations are linked to holes in the neutral interstellar medium.

Finally, Fig.~\ref{ic_mom2} shows the second moment map, or velocity
dispersion of IC~2574. Although some regions near shells reach values
as high as 15 \kms, the overall velocity dispersion in the unperturbed
parts is much lower. Using \ion{H}{1} profiles, we derive the velocity
dispersion in the disk by looking at quiescent regions. Several lines
of sight were averaged to give a one sigma velocity dispersion of
$7\pm 1$\kms\ (similar to what is found in Ho~II). This shows that our
velocity resolution of 2.5 \kms\ (FWHM) was quite sufficient to
resolve the lines of neutral hydrogen.

\placefigure{ic_mom0}

\placefigure{ic_mom1}

\placefigure{ic_mom2}

In this paper, we focus on the small--scale structure of the ISM in
IC~2574. For our analysis, however, we need to adopt values for the
orientation of the object which are usually derived on the basis of a
kinematical analysis (rotation curve fitting).  A full analysis of the
dynamics of IC~2574 based on our observations, however, is beyond the
scope of this paper. Luckily, an adequate analysis has already been
published (see the paper by \markcite{MAR94}MCR94). In the course of
their study they derive a systemic velocity of $58\pm 7$ \kms, adopt
an inclination of $75^{\circ} \pm 7^{\circ}$ and a position angle of
$52^{\circ}$. Depending on their dynamical model, they find that
$70-90\%$ of the total mass of the galaxy is in the form of dark
matter. Following the same procedure as performed for the dwarf galaxy
II~Zw~33 (Walter \ea \markcite{WAL97}1997) we carried through a first
analysis of our data confirming the results by MCR94 regarding the
rotation curve and the global orientation of IC~2574. For ease of
comparison we therefore adopted their values throughout this paper.

\subsection{The scale height of the \ion{H}{1} disk}
\label{scale}

The scale height $h$ of the \ion{H}{1} layer of IC~2574 plays an
important role in the determination of the properties of the holes,
e.g., when translating the observed \ion{H}{1} surface density into an
\ion{H}{1} volume density. Here $h$ is defined as the $1\sigma$ value
for a Gaussian distribution of the \ion{H}{1} layer. Recent studies
have shown that in general dwarf galaxies have thicker \ion{H}{1}
disks than more massive spiral galaxies. Their aspect ratio tends to
be of the order of 10:1 rather than 100:1, e.g., Ho~II: $h=625$ pc
(PWBR92) or NGC~5023: $h=460$ pc (Bottema \ea \markcite{BOT86}1986;
see also Staveley--Smith \ea\ \markcite{STA92}1992). In the case of
IC~2574 there are also indications that the \ion{H}{1} is distributed
in a thicker disk. MCR94 already noticed that, looking at the apparent
axial ratio of the integrated \ion{H}{1} map and assuming a real
inclination of $75^{\circ}$, the \ion{H}{1} must be distributed in a
somewhat thicker disk. Moreover, the fact that we observe, as we will
argue below, spherically expanding shells with diameters of up to almost
1 kpc implies that the extent of the \ion{H}{1} layer must be of the
same order of magnitude. Hence, the scale height of the \ion{H}{1} gas
should be of order 350--400 pc.

An approximate value for the scale height can be obtained fairly
simply, as shown by PWBR92. We follow their approach here. The scale
height is proportional to the velocity dispersion of the gas and
inversely proportional to the square root of the mass density in the
disk (Kellmann \markcite{KEL72}1972; van der Kruit
\markcite{KRU81}1981). For a density distribution that can be
described as $\rho(z,R)=\rho(0,R)\,\mbox{sech}^2(z/z_0)$, $z_0$ is
given by
$$
z_0(R)=\frac{\sigma_{gas}}{\sqrt{2\pi\,G\,\rho(0,R)}}
$$ where $R$
is the galactocentric distance, $\sigma_{gas}$ the velocity dispersion
of the gas in quiescent regions of the galaxy (Sec.~\ref{global}), and
$\rho(R)$ the total volume density (gas, stars and dark matter) of the
galaxy in the plane.  However, since $\mbox{sech}^2(z/z_0)\approx
\mbox{exp}(-z^2/z_0^2)$ (to within 5$\%$ in the region of interest) we
can also assume a Gaussian distribution in $z$. The $1\sigma$ scale
height $h$ is then defined as $h = z_0 / \sqrt{2}$.  To calculate the
average density of IC~2574 in the plane we assume that the total mass
of the galaxy can be estimated from the last measured point
($R_{max}$) of the observed rotation curve (MCR94) and that this is
distributed over a spherical volume of radius $R_{max}$. MCR94 derive
a rotation curve with $v(R_{max})=66$ \kms\ at $R_{max}=8.0$ kpc. This
corresponds to a total dynamical mass of $8.3 \times 10^9$
M$_{\odot}$. Distributing this mass over a spherical volume with
radius $R_{max}$ gives an average value for $\rho$ of 0.0039
M$_{\odot}$ pc$^{-3}$.  Since this is only the mean density of the
galaxy and the density near the plane is certainly higher, we take
twice this derived value (see PWBR92 for a justification) to get a
first approximation to the real volume density $\rho(R)$ near the
center. Using this approach, we find $\rho(R)=0.008$ M$_{\odot}$
pc$^{-3}$.

Substituting the observed velocity dispersion of 7 \kms\ (see
Sec.~\ref{global}) in the equation given above we derive a scale
height of $h=345$ pc for IC~2574. Note that this value, given all the
uncertainties in the derivation, is in good agreement with the value
derived based on the sizes of the largest holes ($h=350-400$ pc). We
therefore adopted a scale height of $h=350$ pc for the \ion{H}{1}
layer of IC~2574.

We can now calculate the volume density of the \ion{H}{1} from the
projected and corrected to face--on \ion{H}{1} surface density. The
surface density profile was obtained by elliptically averaging the
flux-- and primary--beam corrected total \ion{H}{1} map (assuming an
inclination of 75$^{\circ}$). Again, this is correct to first order
only as in the case of a thick \ion{H}{1} distribution the
deprojection of the observed projected \ion{H}{1} distribution is not
trivial. We then calculated the volume density $n_{HI}(R)$ of
the \ion{H}{1} in the plane of the disk using
$$ 
N_{HI}(R)=\int_{-\infty}^{+\infty}n_{HI}(R) \, \exp{(\frac{-z^2}{2
h^2})}\,\mbox{dz} = \sqrt{2\pi} \, h \, n_{HI}(R)
$$

\section{\ion{H}{1} Holes in IC~2574}
\label{hihol}

\subsection{The \ion{H}{1}--hole catalogue}
\label{hicat}

We conducted a search for \ion{H}{1} structures in IC~2574 using the
tasks {\sc kview, kpvslice} and {\sc kshell} that are built into the
recently developed {\sc karma}\footnote{The {\sc karma} visualization
software package was developed by Richard Gooch of the Australia
Telescope National Facility (ATNF)} visualization software
package. {\sc kview} offers many different ways to look at a movie of
a datacube whereas {\sc kpvslice} produces real--time position
velocity (pV) cuts through a cube in any orientation which
tremendously facilitates the search for and identification of
hole--like structures. Finally, {\sc kshell} was used to make
radius--velocity plots of some of the more prominent shells and holes.

First, the {\sc robust} cube was searched for the smaller holes in
IC~2574. In order to be sensitive to more extended structures and
therefore bigger holes, we smoothed our {\sc robust} cube to 11$''$,
15$''$ and 30$''$, respectively, and inspected each of them. For each
\ion{H}{1} hole in IC~2574 we determined its position, size and
expansion velocity. Both authors independetly produced a list of holes
to reduce personal bias, based on some simple criteria regarding the
determination of the center, size, expansion velocity and systemic
velocity of the holes. After that, the two lists were compared and
merged. There was an overlap of about 75$\%$ in terms of well defined
holes which lends some credibility to the identification process.  We
reexamined the remaining structures and, erring on the side of
caution, ended up with a catalogue of 48 holes. Although this list
cannot be free from personal bias, the method closely
follows that employed in the case of M~31 and Ho~II which should make
a comparison between those objects and IC~2574 valid.

It is interesting, at this point, to reflect on the reality of the
features. Do they really constitute large--scale, coherent structures
or are we tricked by nature into seeing coherence in what is basically
a turbulent, perhaps fractal medium? Mac~Low \ea
(\markcite{MAC98}1998) recently produced sets of model data cubes,
similar to those obtained via radio (\ion{H}{1}) or optical
(H$\alpha$) observations (i.e., position, position and velocity). We
analysed, using the same techniques as explained above, a cube of
uniform, isotropic, isothermal, supersonic, super--Alfv\'enic,
decaying turbulence. The simulations assume that the gas is optically
thin and emits with a strength directly proportional to its
density. The model corresponds to a weakly magnetized ISM. Single cuts
through the data cube show features remarkably similar to the ones
seen in the \ion{H}{1} data. However, when focussing on a potential
feature and investigating its morphology in position--position or
position--velocity space, no coherent structure emerges. This lends us
to believe that shells due to random (or at least turbulent) processes
make up a small fraction of the holes we classified. Instead, it is
more likely that the \ion{H}{1} holes and shells are due to mechanisms
which act on large scales, such as stellar winds and (multiple)
supernovae.

Following Brinks \& Bajaja \markcite{BRI86}(1986), we distinguished
between three different types of holes based on their appearance in
the pV--diagrams. A Type 1 hole corresponds to a total blowout (i.e.,
we don't observe the receding nor the approaching side of the
shell). In that case, it is impossible to determine the expansion
velocity and hence the kinematics of the hole. Figure~\ref{pvfinal}c shows a
pV--diagram of a typical Type~1 hole (our hole number 30) using {\sc
kpvslice}.

A Type~2 hole corresponds to a hole which is offset with respect to
the plane of the galaxy. In that case, we witness a deformation in
the pV--diagram towards the plane of the galaxy and a blowout to the
opposite side (see Silich \ea \markcite{SIL96}1996 for numerical
simulations of such a case --- their case C).  The amount of
deformation yields the expansion velocity of the shell towards the
midplane. The part of the shell that points away from the plane is
usually not visible due to lack of sensitivity. Figure~\ref{pvfinal}d
shows a pV--diagram of a typical Type~2 hole (our hole number 33).

Type~3 is the 'classic hole' and corresponds to a more or less
complete expanding shell. The signature in pV--space is that of an
elliptical hole. The expansion velocity is equal to half the
difference between the velocity of the approaching and receding
sides. The systemic velocity of the hole is defined by its
center. Figure~\ref{pvfinal}a shows a pV--diagram of a typical example of
a Type~3 hole (our hole number 21). Figure~\ref{pvfinal}b shows the same
hole, but analyzed with the task {\sc kshell}. In this figure, the
horizontal axis corresponds to radius $r$ (distance from the center of
the hole) and the intensity corresponds to the average \ion{H}{1}
density within an annulus with radius $r$ around the center of the
hole. The vertical axis is again velocity. Since we are looking at a
radius--velocity diagram, we will call such a representation an
rV--diagram in the following. For an ideal shell, a half ellipse
should be visible in which the diameter in the $v$--direction is twice
the expansion velocity and its dimension in the $r$--direction the
radius of the shell. The advantage of looking at a near circular
\ion{H}{1} shell with {\sc kshell} lies in a substantial increase in
signal--to--noise ratio as one averages over many pixels whereas in a
pV--diagram one merely cuts through an \ion{H}{1} structure.

Note that in the case of a Type~1 hole, the absence of detectable
emission corresponding to the front and back of a shell causes a hole
in rV--space. For obvious reasons, a Type~2 hole does not show a
prominent signature in an rV--diagram either.

\placefigure{pvfinal}

As was already mentioned earlier, the holes were found to be much more
prominent in the channel maps than in the integrated \ion{H}{1}
map. This is due to the high inclination of IC~2574 which leads
to several \ion{H}{1} shells at different locations within the disk
and possibly different velocities overlapping along the
line--of--sight, as is the case in M~31.  The total \ion{H}{1}
surface brightness map was therefore of little use for the
identification of a hole. Note that in the case of Ho~II, {\em only}
the integrated map was used because of the much lower inclination of
the galaxy. This implies that the approach to search for holes needs
to be adapted to the orientation of the galaxy.

Figure~\ref{ic_holes} shows a plot of the distribution of the 48
\ion{H}{1} holes detected with confidence in IC~2574. The greyscale
map is a linear representation of the total \ion{H}{1} surface
brightness. Some of the holes are clearly visible in the total
\ion{H}{1} map but the majority of holes doesn't show up at all and
can only be detected in the channel maps or pV--diagrams.

\placefigure{ic_holes}


The final list of the detected \ion{H}{1} holes is given in
Table~\ref{holecat}. The properties listed are as follows:

\begin{description}

\item[Column 1:] Number of the hole (increasing with increasing right
ascension)

\item[Column 2, 3:] Position of the center of the hole in right
ascension and declination, respectively (B1950.0 coordinates). Typical
errors are 3$''$.

\item[Column 4:] Heliocentric velocity $v_{\rm hel}$ of the hole defined
by the channel in which the hole is largest or most prominent. The
accuracy of $v_{\rm hel}$ is about 2.5 \kms\ (the velocity resolution of
our observations).

\item[Column 5:] Diameter $d$ of the hole in kpc. Note that most of
the holes are circular. In the case of elliptical holes, the diameter
along the major axis, along the minor axis and the position angle
($2 \times b_{\rm maj}, 2 \times b_{\rm min}, PA$) is given 
(but see Sec.~\ref{single}). The accuracy of
$d$ is around 95 pc (the FWHM of our synthesized beam).

\item[Column 6:] The expansion velocity $v_{\rm exp}$ of the hole in
\kms. Note that in the case of a Type 1 hole, no expansion velocity
can be measured. Typical errors are about 2 \kms.

\item[Column 7:] The estimated \ion{H}{1} volume density $n_{\rm HI}$ (in
M$_{\odot}\,$pc$^{-3}$) of the region where the hole is situated before
the hole was created. $n_{\rm HI}$ was calculated based on the
\ion{H}{1} surface density, as described in Sec.~\ref{scale}.
Given the uncertainties in the deprojection and the assumptions which
went into the derivation of the mass density within the disk, we
estimate a relative error in the value of $n_{\rm HI}$ to be of order
30$\%$.

\item[Column 8:] The type of the hole (1--3), as defined in
Sec.~\ref{hicat}.

\end{description}

Table~\ref{propcat} summarizes the physical properties derived from
Table~\ref{holecat}. The properties listed are as follows:

\begin{description}

\item[Column 1:] Number of the hole

\item[Column 2:] Diameter $d$ of the hole. For elliptical holes, the
diameter was defined using the geometric mean 
$$
d=\sqrt{2 \, b_{\rm maj} \times 2 \, b_{\rm min}}
$$

\item[Column 3:] Expansion velocity $v_{\rm exp}$ of the hole in
\kms. Same as column 6 in Table~\ref{holecat}.

\item[Column 4:] Kinematic age of the hole in units of $10^6$
years. This age determination assumes that over its entire lifetime
the hole was expanding with $v_{\rm exp}$. Hence:
$$
t=\frac{(d/2)}{v_{\rm exp}}
$$ 
The accuracy of the kinetic age is
typically 20$\%$. Because of the assumption of a constant expansion
velocity, the kinematic age is an upper limit.

\item[Column 5:] Indicative \ion{H}{1} mass $M_{\rm HI}$ in units of
$10^4$ M$_{\odot}$. It is the mass which filled the observed hole
under the following assumptions: i) the hole is spherical, ii) the gas
was distributed uniformly in the galaxy, and iii) the hole is
completely empty now. Note that if one can assume that only a minor
fraction of the surrounding \ion{H}{1} is ionized, this mass corresponds
also to the total \ion{H}{1} mass on the rim of the hole. $M_{\rm HI}$ was
calculated using
$$
M_{\rm HI}=\frac{4}{3}\,\pi \, (d/2)^3 \, n_{\rm HI}(R)
$$ where
$n(R)$ is the volume density of the gas at the galactocentric distance
$R$ (column 7 in Table~\ref{holecat}). To correct this value for the
amount of primordial helium, a factor of 1.33 has to be applied. The
relative error is, as $d$ enters to the third power, of order
30$\%$. Note that in the case of the largest holes, this value is more
likely to be an upper limit since $n_{\rm HI}$ decreases with
increasing distance $z$ from the plane.

\item[Column 6:] The initial total energy $E_{\rm E}$ that was
released by the supernova explosions which created the hole (in units
of $10^{50}$ ergs).  Note that the kinetic energy of the expanding
shell is about 1$\%$ to $10\%$ of $E_{\rm E}$ (depending on the
ambient density, see Chevalier \markcite{CHE74}1974).  To determine
$E_{\rm E}$, we used the equation by Chevalier \markcite{CHE74}(1974)
who derived a fit to numerical results of an expanding shell created
by a single explosion (see also McCray \& Snow \markcite{MAS79}1979).
Expressing $E_{\rm E}$ in terms of observable quantities $d$, $v_{\rm
exp}$ and $n_0$, he obtained
$$
E_{\rm E}=5.3\times 10^{43}\, n_0^{1.12}\, (d/2)^{3.12}\, v_{\rm exp}^{1.4}\,
\mbox{ergs} 
$$ 
where $v_{\rm exp}$ is the expansion velocity in \kms, $n_0$ the
density of the ambient medium in particles per cubic centimeter, and
$d$ is the diameter of the shell in parsecs. In our calculations we
replaced $n_0$ by $n_{\rm HI}$ as was done, e.g., for M~31 and
Ho~II. This is not strictly correct as we are ignoring contributions
by He and heavier elements. To correct $E_{\rm E}$ for this
contribution, all values should be multiplied by a factor of $\approx
1.5$. In addition, the derived \ion{H}{1} densities are average values
for a given galactocentric distance which introduces another source of
uncertainty. Note that the numerical coefficient in Chevaliers
equation depends on the radiative energy loss rate and thus on the
heavy element abundance in the expanding shell. Not only is this
abundance different for various galaxies, but it decreases with
galactic radius, both for our own and external galaxies and by more
than an order of magnitude (e.g., Heiles \markcite{HEI79}1979). The
last phase in the expansion of a shell is reached when, after the
shell has continued to collect more of the ambient medium it will slow
down to velocities comparable to the random motions of the ISM (around
7 \kms) and hence stops expanding. Most probably the largest
structures we observe are in this latter phase (see
Sec.~\ref{large}). From this brief discussion it should be clear that
the derived energies $E_{\rm E}$ are only order of magnitude
estimates.

\placetable{holecat}

\placetable{propcat}

\end{description}

\subsection{Notes on single holes}
\label{single}

The following holes deserve a more detailed description:

\begin{description}

\item[Hole $\#$35:] This hole causes a major disturbance in the
pV--diagrams of IC~2574. It is one of the few elliptical holes, with
dimensions of $1000 \times 500$ pc. Although there is some
circumstantial evidence that it is a superposition of two spherical
holes, it was treated as one hole. As we will show in
Sec.~\ref{ha-hi}, the most prominent \ion{H}{2} region in IC~2574 is
located on the rim of this supergiant shell. The shell is filled with
soft X--ray emission as observed with the ROSAT telescope. A detailed
multiwavelength analysis (optical, radio, X--ray) of this particular
supergiant shell has been presented elsewhere (Walter \ea
\markcite{WAL98}1998)

\item[Hole $\#$22:] Catalogued elliptical in shape, this hole is
actually a superposition of several holes, leading to the overall
impression of an elliptical feature; we treated this as a single
object since subdividing it in, say, three holes would have been
arbitrary.

\item[Holes $\#$ 13, 23, 26:] As can be seen from the channel maps
(Fig.~\ref{ic_cm_1}), some holes are not clear--cut cases in terms of
the definition of a hole given in Sec.~\ref{hicat}.  They are possibly
chance superposition of old shells and fragments leading to the
impression of a hole in the integrated \ion{H}{1} map (see also
Sec.~\ref{large}). We therefore decided not to include these holes in
the statistical analysis.

\end{description}

\subsection{Notes on large--scale structures}
\label{large}

As can be seen from Table~\ref{holecat}, only few shells larger then
$\approx 1000$~pc have been catalogued in IC~2574. It has to be
mentioned, however, that there appear coherent features larger then
1~kpc in the channel maps. Several of these can be identified in the
channel maps (Fig.~\ref{ic_cm_1}), e.g., between 45 and 70 \kms. Note
that these structures are related to the major disturbance in the
velocity field due north of the center of the galaxy. Obviously, it is
very difficult to catalogue these features as they lack the clear
coherence, both in position as well as in velocity, of \ion{H}{1}
shells. Moreover, because of their large size, their dynamics is not
dominated by some leftover expansion but rather by large scale
galactic rotation modified by the effects of nearby, smaller
holes. The reason why we wish to mention these large scale features is
because along some of the apparent arcs we find \ion{H}{2} regions.

\section{The \ion{H}{1}--H$\alpha$ connection in IC~2574}
\label{ha-hi}

\subsection{Global correlations}

Until now, we have focussed on the structure and dynamics of the
neutral ISM of IC~2574. In the following, we will compare these
results with our H$\alpha$ observations which reveal the sites of
recent or ongoing SF in IC~2574.  As noted earlier (Sec.~\ref{Intro}),
the expanding \ion{H}{1} structures are widely thought to result from
recent star formation. They sweep up gas causing the volume density on
the rim of the shell to increase until gravitational instabilities
lead to molecular cloud formation and hence secondary SF (e.g.,
Palou\v s \ea \markcite{PAL90}1990, Elmegreen
\markcite{ELM94}1994). Therefore, a comparison of the position of
\ion{H}{1} holes with \ion{H}{2} regions should show some correlation,
the H$\alpha$ emission being predominantly found along the rims of
larger holes, as evidence for propagating SF. In the case of the
smaller, young holes one would expect some to be still filled with
diffuse H$\alpha$ emission coming from the ionized wind--blown
cavities around the still present, young O and B stars.

Figure~\ref{haho} shows an overlay of the H$\alpha$ map with the
\ion{H}{1} holes we detected in our survey (Sec.~\ref{hicat}). The
width of the rims of the \ion{H}{1} holes is of course in reality
larger then indicated in this plot. The subregions A--C will be
discussed in more detail in the following section. As can be seen from
this overlay, there is a prominent correlation of ongoing SF with the
rims of the large \ion{H}{1} shells which is in general agreement with
the picture painted above.  There is a strong tendency for \ion{H}{2}
regions to be situated on the rims of the \ion{H}{1} shells (for some
very nice examples see holes 8 and 35) and only in a very few cases
diffuse H$\alpha$ emission does not seem to be related to an
\ion{H}{1} structure. Note that in one case only (hole 32), diffuse
H$\alpha$ emission (partly) fills an \ion{H}{1} hole.

Fig.~\ref{hahic} presents an overlay of the H$\alpha$ map with one
contour of the \ion{H}{1} surface brightness map (flux and primary
beam corrected) plotted at $1.7\times 10^{21}$ atoms cm$^{-2}$. As can
be seen from this overlay, H$\alpha$ emission is globally correlated
with a high \ion{H}{1} surface brightness (for an exception in the
very north see below). Kennicutt \markcite{KEN89}(1989) used the disk
instability relation of Toomre \markcite{TOO64}(1964) to show that,
for spirals, there is a threshold density for star formation. For gas
densities above the threshold, the star formation rate (SFR) is
consistent with a Schmidt law. Near the threshold, the correlation
breaks down, and one sees bursts of SF. Much below the threshold the
SFR is extremely low. Skillman \markcite{SKI87}(1987) has
observationally determined this threshold for dwarf galaxies to be on
the order of $1\times 10^{21}$ atoms cm$^{-2}$ (at a resolution of
500~pc) which is in general agreement with our results (see also
below). For a detailed discussion on the models that describe the
relationship between gas, stars and star formation in irregular
galaxies, the reader is referred to the paper by Hunter, Elmegreen \&
Baker \markcite{HUN98}(1998).

\placefigure{haho}

\placefigure{hahic}

\subsection{Analysis of individual \ion{H}{2} regions}
\label{sha}

Using our high resolution \ion{H}{1} data of IC~2574 we are now able
to perform a much more detailed analysis and comparison of the
properties of the \ion{H}{2} regions with the \ion{H}{1} features that
are related to them. When making overlays of the H$\alpha$ map with
the high resolution cube we found that in virtually all cases, single
\ion{H}{2} regions are correlated with distinct \ion{H}{1} structures
like clumps or clouds. We defined 40 regions where H$\alpha$ emission
and \ion{H}{1} clouds seem to be related. These regions are outlined
by the squares in Figs.~\ref{har1}--\ref{har3} which are meant to
indicate which \ion{H}{2} regions were lumped together.  They are
numbered 1$'$ to 40$'$ to avoid confusion with the numbers of the
\ion{H}{1} holes.  Note, that sometimes two or more H$\alpha$ features
belong to the same \ion{H}{1} structure (like a clump) and were thus
considered as one region (e.g., in region 40$'$), whereas in other
cases individual \ion{H}{2} regions were treated as separate entities
(e.g., regions 11$'$ and 12$'$). The H$\alpha$ fluxes of the regions
tell us something about the current SF rate and the average number
density of the ionized gas whereas the \ion{H}{1} data yield
approximations for the number density of neutral hydrogen, the
velocity dispersion of the gas in that region and the thresholds for
SF. Figs.~\ref{har1}--\ref{har3} are enlargements of the regions A--C
marked in Fig.~\ref{haho} and display the regions defined by us. The
position and size of the \ion{H}{1} holes are plotted as well for
comparison (please refer to Fig.~\ref{haho} for cross referencing of
the holes). The fluxes of the \ion{H}{2} regions were taken from the
observations of \markcite{MIL94}MH94.  Table~\ref{cross} summarizes
the cross identifications between their and our regions. In this table
we list the coordinates of the centers of our regions as well.

Table~\ref{Ha-regions} summarizes the results of our analysis of the
\ion{H}{2} regions. The first column refers to the number of the
\ion{H}{2} region as defined in Figs.~\ref{har1}--\ref{har3}. From the
fluxes published by \markcite{MIL94}MH94, we deduced the total flux
$F(H\alpha$) of our groups by simply summing their fluxes in the
respective regions (column~2, Table~\ref{Ha-regions}). From this
number, the H$\alpha$ luminosity can be calculated using
$L(H\alpha)=4\pi\,D^2 F(H\alpha),$ which, in the case of IC~2574
($D$=3.2~Mpc), translates to $L(H\alpha)=3.1\times 10^{17} F({\rm
H\alpha})$ where $F(H\alpha)$ is given in ergs~cm$^{-2}$~s$^{-1}$ and
$L(H\alpha)$ in units of solar luminosities $L_{\odot}$ ($1
L_{\odot}=3.85\times 10^{33}$ ergs~s$^{-1}$). The number of Lyman
continuum photons per second is given by $N_{\rm Lyc}=2.82\times
10^{45} L({\rm H\alpha}).$ Table~\ref{sprop} (see Devereux \ea
\markcite{DEV97}1997 and references therein) summarizes the expected
properties of high mass stars; column~1 gives the spectral type,
column~2 the stellar mass, column~3 the bolometric stellar luminosity,
column~4 the number of Lyman continuum photons per second, column~5
the H$\alpha$ luminosity L$(H\alpha)$, and column~6 the ratio of
bolometric to H$\alpha$ luminosity. These equations and values can be
used to estimate, for example, the equivalent number of O5 stars, for
each region (see below).

Column 3 in Table~\ref{Ha-regions} presents the volume density of the
ionized free electrons over the area where we find H$\alpha$ emission
(as defined by MH94). This was arrived at by first calculating
the emission measure (EM) of each of their regions according to:
$$
EM=2.41\times 10^3 \, T^{0.92}\, F(H\alpha)\, {\rm cm}^{-6}\, {\rm pc}
$$
(Peimbert \ea \markcite{PEI75}1975), where $F(H\alpha)$ is 
the H$\alpha$ flux in units of ergs~cm$^{-2}$~s$^{-1}$~sterad$^{-1}$ 
and $T$ is assumed to be the canonical value of $10^4$~K. 
From that, the mean electron density for every 
single region published by MH94 was derived using
$$
EM=n_{\rm e}^2\, d\, {\rm cm}^{-6}\, {\rm pc}
$$
where $n_{\rm e}$ is the electron density and $d$ the path length in
pc through the \ion{H}{2} region (assuming a spherical distribution of
the H$\alpha$ emission). Since the derived values for $n_{\rm e}$ were
very similar for a single region, we calculated the value for the
electron density for each of our regions by taking an
intensity--weighted average of the contributing MH94 regions. The
average electron densities range from 0.5 to 2 cm$^{-3}$. This is,
most likely, a lower limit for the actual electron density due to
internal absorption within IC~2574. Note that in the above, effects
due to the filling factor of the gas are ignored. The real densities
in the \ion{H}{2} regions are likely much higher.

In order to derive properties of the \ion{H}{1} features that are
physically related to the \ion{H}{2} regions, we defined areas $A$ around
each of the \ion{H}{2} regions with an effective radius given by $r_{\rm
eff}=\sqrt{A/\pi}$, where $r_{\rm eff}$ is given in pc (column
4). Although this is a somewhat subjective process, we found that slight
variations in the size of the aperture didn't change the derived
properties by more than $20-30\%$. Column 5 represents the total
\ion{H}{1} mass found within the aperture using the flux and primary beam
corrected \ion{H}{1} data. For every single region, we measured the
velocity dispersion of the \ion{H}{1} by taking spectra through the
\ion{H}{1} cube.  Column 6 summarizes the measured velocity widths
defined as the full width at half maximum (FWHM) from which the real
velocity dispersion $\sigma_v$ (assuming a Gaussian velocity
distribution) can be calculated using
$
\Delta v=2\sqrt{2\ln 2}\,\sigma_{\rm v}=2.355\, \sigma_{\rm v}
$ 
Interestingly, these values tend to be high where two or more shells 
overlap (e.g., the \ion{H}{2} complexes 16, 31, 32, 33). These high 
dispersions are probably caused by blending along the line of sight of
these shells.

Column 7 presents the average density of the neutral hydrogen which
was calculated using
$$
\rho_{\rm HI}=M_{\rm HI}/(\frac{4}{3}\, \pi\, r_{\rm eff}^3)
$$
with $M_{\rm HI}$ and r$_{\rm eff}$ the values from columns 5 and 4,
respectively.  Values for $\rho_{\rm HI}$ range from about 1 cm$^{-3}$
to values as high as 6 cm$^{-3}$. Note that these are only averages
and that locally the \ion{H}{1} density might be more
enhanced. Based on the analysis by MCR94 and some crude mass modeling
(Sec.~\ref{scale}) we derive an average density in the disk of
0.008 $\rm M_\odot\,pc^{-3}=0.3\, atom\,cm^{-3}$. The regions in which
star formation is currently active thus have an overdensity
of an order of magnitude (see Table~\ref{Ha-regions}).

The average surface density of the \ion{H}{1} regions is shown in
column 8. Note that virtually all values are larger than $1\times
10^{21}$~cm$^{-2}$ which is in excellent agreement with the
observational value given by Skillman \markcite{SKI87}(1987).  In
other words, SF is taking place if the projected \ion{H}{1} surface
density reaches values of about $10^{21}$ cm$^{-2}$ not only for a
resolution of 500 pc, as observed by Skillman but also for almost an
order of magnitude improvement ($\approx 95\, \rm pc$)! The highest
surface density found is of order $2.5\times 10^{21}$~cm$^{-2}$.

Finally Column 9 summarizes some remarks regarding the regions. Most
of the \ion{H}{2} regions are related to the rim of an \ion{H}{1}
shell. It should be noted at this point that in the course of checking
for correlations between H$\alpha$ and \ion{H}{1} emission, we
detected 5 new tiny holes at the positions were there is H$\alpha$
emission. However, as these holes were not detected in our
inspection of the \ion{H}{1} cubes we didn't include them in the final
catalogue which is thought to be complete for holes with radii larger
then about 100 pc only (as discussed in Secs.~\ref{hicat}
and~\ref{stat}).

We catalogued four of the \ion{H}{1} features as ``clumps'' (regions
1$'$, 2$'$, 6$'$ \& 22$'$). By 'clump' we mean a well--defined
\ion{H}{1} region in 2 spatial dimensions as well as in velocity
space.  These clumps look like small, filled circles in
position--velocity space. For those cases we asked ourselves whether
these clumps are gravitationally confined or not. To answer that
question, we calculated the virial masses of these objects using
$$
M_{\rm vir}=250\, \Delta v^2 \, r_{\rm eff} \, {\rm M}_\odot
$$
(Rohlfs \& Wilson \markcite{ROH96}1996), where $\Delta v$ is the full
velocity width 
at half maximum (FWHM) in km~s$^{-1}$ and $r_{\rm eff}$ the radius of
the clump in pc. The virial masses of the four regions are: $M_{\rm
vir}(1$'$) = 13\times 10^6 \,{\rm M}_\odot$, $M_{\rm vir}(2$'$) =
38\times 10^6 \, {\rm M}_\odot$, $M_{\rm vir}(6$'$) = 21\times 10^6 \,
{\rm M}_\odot$ and $M_{\rm vir}(22$'$) = 7\times 10^6 \, {\rm
M}_\odot$. As we see, the virial masses are all about one order of
magnitude higher than the \ion{H}{1} masses. So, the clumps are either
not gravitationally bound or they are held together by as yet
undetected material. One possibility is that they contain an order of
magnitude more mass in the form of molecules. This will be hard to
confirm, however, as dwarf galaxies have been notoriously difficult to
detect in CO, the most abundant tracer for molecular gas (e.g., Sage \ea
\markcite{SAG92}1992). Dark Matter could play a role, but the mass
density, including DM, near the above mentioned SF regions is of order
0.008~$M_\odot\,{\rm pc^{-3}}$ (see Sec.~\ref{scale}), insufficient by
a wide margin.

If the observed velocity dispersion is not due to gravitation, the
question remains, why {\em do} we see clumps? We can think of two more
mechanisms, magnetic confinement or pressure. The former is unlikely
to play a large role. Dwarf galaxies do not possess large scale
magnetic fields (the exceptions being NGC~4449 and IC~10, e.g., Chyzy \ea
\markcite{CHY98}1998), any non--thermal radio continuum radiation being confined to the
strongest regions of active star formation (Deeg \ea
\markcite{DEE93}1993).

We did not find any correlation of H$\alpha$ flux with the surrounding
\ion{H}{1} density. However, one wouldn't really expect any strong
correlation of this kind. Not only is the H$\alpha$ emissivity of a
star forming region strongly dependent on the age of the region (see,
e.g., Gerritsen \& Icke 1997), but it is also dangerous to try to
deduce actual physical properties of the SF regions from mean values,
like average electron or \ion{H}{1} densities, while ignoring
spatial variations in their volume filling factors.

In some regions within IC~2574 (e.g., region 40'), we find high
\ion{H}{1} surface densities but only little H$\alpha$ emission
indicating that possibly a major star formation burst is about to
commence there and that therefore most of the gas is still
neutral. With the help of the equations given above we can calculate
how many stars are approximately needed to create the observed
H$\alpha$ flux.  In the case of region 40$'$ we find a flux of
$1.5\times 10^{-14}\, \rm ergs\, cm^{-2}\, s^{-1}$ and deduce a
luminosity of $4.7\times 10^3\, {\rm L}_\odot$ which is the equivalent
of a few O7 stars only (see Table~\ref{sprop}).  Note, however, that
there exist also regions where the situation is reversed and most of
the \ion{H}{1} is being ionized by the Lyman continuum photons of the
most massive stars of the powerful \ion{H}{2} regions. Region 34' is a
prominent example (see also Fig.~\ref{hahic}). From the H$\alpha$ flux
of $F(H\alpha)\approx 1\times 10^{-12}\, \rm ergs\, cm^{-2}\, s^{-1}$
we derive an H$\alpha$ luminosity of $L(H\alpha)=3.1\times10^5\, {\rm
L}_\odot$ which corresponds to an equivalent of 22 O5 stars, in other
words a major SF region.

\placetable{cross}

\placetable{Ha-regions}

\placetable{sprop}

\placefigure{har1}

\placefigure{har2}

\placefigure{har3}

\section{Discussion}
\label{discu}

\subsection{What created the \ion{H}{1} holes?}
\label{what}

The ISM of IC~2574 is dominated by expanding \ion{H}{1} holes and
shells. Similar structures have been found in other gas--rich dwarf
galaxies (e.g., Holmberg~II (PWBR92), IC~10 (Wilcots \& Miller
\markcite{WIL98}1998) and in more massive spiral galaxies like our
Galaxy (Heiles \markcite{HEI79}1979), M~31 (Brinks \& Bajaja
\markcite{BRI88}1988) and M~33 (Deul \& den Hartog
\markcite{DEU90}1990). One of the major questions concerning these
structures is what physical process or processes are responsible for
their origin.

Heiles (\markcite{HEI79}1979), who was the first to recognize
expanding \ion{H}{1} shells in the Galaxy, concluded that 'many of the
\ion{H}{1} structures [...]  were probably produced by large amounts
of energy injected directly to the interstellar medium'. One obvious
mechanism is the after effects of massive star formation: strong
stellar winds and supernova explosions (see the review by
Tenorio--Tagle and Bodenheimer 1988). Detailed numerical modeling by
various authors (e.g., Palou\v{s}, Franco \& Tenorio--Tagle
\markcite{PAL90}1990, Silich \ea \markcite{SIL96}1996, Palou\v{s},
Tenorio--Tagle \& Franco \markcite{PAL94}1994) has strengthened
confidence in this explanation, as has the work by Oey \& Clarke
\markcite{OEY97}(1997) who in a (semi--)analytical way managed to
predict the \ion{H}{1} shell size distribution in the SMC based on the
observed H$\alpha$ luminosity function.

Having said that, there are potentially problems with this
interpretation, as mentioned by Radice, Salzer \& Westpfahl
(\markcite{RAD95}1995) and Rhode, Salzer \& Westpfahl
(\markcite{RHO99}1999). Looking for remnant stellar clusters at the
centers of several large holes in Holmberg~II they failed to find a
population of A and F main sequence stars. Moreover, it is difficult
to imagine that the largest \ion{H}{1} shells which are located beyond
the optical galaxy should have been created by star formation (an
argument which holds equally well for some of the holes in
IC~2574). This is corroborated by ultraviolet imaging of Holmberg~II
by Stewart (\markcite{STE99}1999) who finds no FUV emission from the
outer regions of Holmberg~II.

It should be noted, though, that the nondetection of population A and
F stars does not necessarily mean that none of the \ion{H}{1} are
created by SF. In dwarf galaxies, in the absence of shear, \ion{H}{1}
shells can reach large dimensions and grow so old that no obvious
bright cluster remnant will remain visible in the center (Efremov,
Elmegreen \& Hodge \markcite{EFR98}). Some of the observed supershells
are so old that the brightest members of stellar cluster which created
the hole may have evolved off the main sequence and dispersed (assuming
a similar IMF as in the Galaxy). This may make a detection of old
clusters difficult. Such a case (LMC~4/Constellation~III) is described
by Efremov \& Elmegreen (\markcite{EFE98}1998).

Another potential problem with the conventional picture for the origin
of the holes is that for the largest ones the energy requirements seem
to surpass the amount that can possibly be released into the ambient
ISM via stellar winds and supernova explosions.  The giant shell in
M~101 (van der Hulst \& Sancisi \markcite{HUL88}1988) is just one
example of such a structure. Tenorio--Tagle (\markcite{TEN87}1987)
proposed as an alternative the infall of high velocity clouds which
could deposit an amount of energy of 10$^{52}$ to 10$^{54}$~ergs per
collision (see the review by Tenorio--Tagle \& Bodenheimer
\markcite{TEN88}1988 for further references). Even though this
mechanism seems a viable explanation for at least some holes, it is
unlikely that it plays a dominant role. Certainly it can't be as
important in dwarf galaxies because of their much smaller collisional
cross section.

Recently, an alternative explanation for the creation of supergiant
\ion{H}{1} shells in galaxies has been proposed, namely GRBs or Gamma
Ray Bursters (Efremov, Elmegreen \& Hodge \markcite{EFR98}1998, Loeb
\& Perna \markcite{LOE98}1998). Even though the basic physics of GRB
are not yet known the picture to emerge is that they are highly
energetic events (of order $10^{54}$~ergs) that release huge amounts
of energy in a very short period of time (order of a few seconds).

Although we recognize that perhaps not all holes can be explained by
stellar winds and supernova explosions, we feel that there is
sufficient evidence for the conventional interpretation to warrant an
analysis in those terms, which is what we pretend in the rest of this
paper.

\subsection{Statistical properties of the holes: a comparison
with other galaxies}
\label{stat}

For the first time we are now in a position to address an important
question regarding the creation and evolution of hole--like structures
in the interstellar medium: how are their properties related to their
environment, i.e., the type of host galaxy. In what follows we will
compare the observed and derived \ion{H}{1} hole properties of IC~2574
with those found in M~31 (Brinks \& Bajaja \markcite{BRI86}1986), an
example of a massive spiral galaxy similar to our own, M~33 (Deul \&
den Hartog \markcite{DEU90}1990), a less massive spiral and Ho~II
(PWBR92\markcite{PUC92}), another dwarf galaxy in the same group of
galaxies, yet four times less massive than IC~2574. In other words,
the sequence M~31 -- M~33 -- IC~2574 -- Ho~II spans a large range of
different Hubble types from massive spirals to low--mass dwarfs. 

Despite the fact that a similar study, and at considerably higher
sensitivity, is now available for the SMC (Staveley--Smith \ea
\markcite{STA97}1997), we decided not to include their results. The
main reason being that the linear scales (or spatial frequencies)
sampled by the ATNF hardly overlap with those observed in the galaxies
listed above. The linear resolution, at 28 pc, is almost four times
higher than, for example, the VLA maps on IC~2574. At the other end of
the spectrum, because of the lack of short spacing information,
structures larger than a few hundred parsec will have been
missed. Moreover, the SMC is a very disturbed system, being torn apart
by tidal forces due to interactions with the LMC and the Galaxy.

Limiting the comparison to the four galaxies listed above has some
further advantages. The linear resolutions are very similar, as are
the velocity resolutions with which they have been observed (see
Tab.~\ref{diff} for a summary). In addition, all four galaxies were
examined in more or less the same fashion, one of the authors (EB)
having taken part in the analysis of three of the four objects.  We
are aware that the results for the four galaxies suffer partially from
low statistics and incompleteness due to personal bias and
observational constraints (such as the beamsize). However, we feel
that these effects, to first order, affect a comparison in a similar
way and that it is valid to try to find global trends as a function
of Hubble type.

For the sake of brevity, rather than first show the statistical
properties of the holes in IC~2574 separately, we immediately go to a
comparison with the other three objects. In
Figs.~\ref{massc}--\ref{velc} we compare the distributions of
swept--up matter, the energies, diameters, expansion velocities and
ages of the holes in the 4 galaxies. In each graph we plot in the form
of histograms the relative number of holes, in percent, in order to
make a direct comparison possible. In each histogram, the data for the
individual objects are binned in the same way. To improve the
presentation, however, the respective bins were slightly shifted when
plotting the results.

Fig.~\ref{massc} shows an overlay of the indicative masses of the
\ion{H}{1} holes, binned logarithmically. The values range from
typically $10^4$ to $5 \times 10^6$ M$_\odot$. Even though a
two--sample Kolmogorov--Smirnov test rejects the hypothesis of
pairwise equality of the distributions, a casual inspection shows that
the indicative masses span roughly the same range.  The mean masses
(in 10$^5$M$_{\odot}$) for the dwarf galaxies (Ho~II: 10.5, IC~2574:
8.4) are slightly higher as compared to the spirals (M~33: 4.1, M~31:
3.2). This is somewhat surprising if one realizes that the volume
density in the plane of a spiral galaxy is typically higher by a
factor of three as compared to that found in a dwarf
galaxy. \ion{H}{1} column densities have a much smaller variation from
object to object, though (see Tab.~\ref{diff}) which probably
explains why the swept up matter works out to be more or less the
same. As a warning, it should be born in mind that, in order to
determine the \ion{H}{1} densities we smeared out any fine scale
structure, potentially adding somewhat to the errors.

Fig.~\ref{energyc} is similar to Fig.~\ref{massc}, showing instead the
distribution of the logarithm of the energies needed to create the
observed \ion{H}{1} holes (under the assumption of them being created
by supernova explosions). The derived energies range from $10^{50}$ to
$10^{53}\, \rm ergs$. We find the important result that the
distributions are identical, at least to first order (mean energies
(10$^{52}$\,ergs): Ho~II: 1.4, IC~2574: 1.1, M~33: 1.6, M~31: 1.5). A
two--sample Kolmogorov--Smirnov test reveals that there is no
significant pairwise difference at the 95\% level between the four
energy distributions. From this we conclude that regardless of Hubble
type of the parent galaxy, the star clusters (or whatever other
process created the \ion{H}{1} holes) deposit more or less the same
amount of energy into the ISM.

Fig.~\ref{diamc} shows an overlay of the size distribution of the
holes found in the four galaxies. In this and the following plots the
bins are on a linear scale. Note that there is a clear difference with
Hubble type! The size distribution for holes in M~31 and M~33 cuts off
sharply near 600 pc. In contrast, holes in IC~2574 and Ho~II reach
sizes of 1200 to 1500 pc, respectively. The lack of holes with sizes
smaller than $\sim 100$~pc is due to our resolution limit. The fact
that the holes get larger for ``later'' Hubble types (i.e., galaxies
with lower mass) can be understood as follows. Smaller galaxies have
lower masses and their mass surface density in the disk is lower as
well. The one--dimensional velocity dispersion varies little, if any,
among galaxies. We find in the four objects under scrutiny, taking
into account an uncertainty of $\pm$ 1\kms, a range of 6--9
\kms. This, combined with the lower mass surface density, leads to
dwarf galaxies having much thicker disks with aspect ratios of 1:10
rather than 1:100. Hence, for the same amount of energy deposited, as
argued above, an \ion{H}{1} shell can grow much larger, because of a
lower gravitational potential and a lower ambient density. Moreover,
in dwarf galaxies shells are prevented from breaking out of the disk
because it is thicker (see Sec.~\ref{scale}).

The age distribution of the holes (Fig.~\ref{agec}) also shows a
sequence with Hubble type. The mean values for the ages of the holes
are: $10 \times 10^6$ years (M~31); $12\times 10^6$ years (M~33); $23
\times 10^6$ years (IC~2574) and $62\times 10^6$ years (Ho~II).  In
the spirals no holes are found which are older than 30 Myr whereas the
distribution in dwarfs is flatter and spans an age range of up to 120
Myr. This is not unexpected since in more massive galaxies shear and
mixing due to the passage of density waves tend to destroy holes. In
dwarfs, unless a shell is hit by a neighboring explosion, and
subsequently expanding structure, a feature can potentially persist
for a long time, until the rim disperses and mixes with the
surrounding ISM.

Finally, the distribution of the expansion velocities is presented in
Fig.~\ref{velc}.  The predominantly low values for the dwarfs
(especially Ho~II) reflect, again, the fact that we see many large
holes in the dwarfs which have almost stalled. A word of caution
should be added as the velocity resolution of the IC~2574 and Ho~II
observations was a factor of about 3 better than those of M~31.  Very
young (i.e., smaller than $\sim 100$ pc) and fast expanding holes are
missed, of course, but this holds true for all galaxies in the sample.
 
Figure~\ref{v_d_comp} displays an alternative way of looking at the
properties of the \ion{H}{1} holes on a galaxy--wide scale. It shows
the expansion velocities of the shells in the four nearby galaxies
plotted as a function of their diameter (crosses). Based on the
hydrodynamical models for expanding shells (Chevalier
\markcite{CHE74}1974), we plotted lines of constant energy for the
expanding structures in Fig.~\ref{v_d_comp} (assuming a constant
density of the ambient medium of 0.15 cm$^{-2}$). The plotted curves
represent energies at a level of $10^{49}$, $10^{50}$, $10^{51}$ and
$10^{52}$ ergs, respectively (from left to right).  The evolution of
an individual hole of a given energy is along an equi--energy line
from the top of the plot downwards. In addition, lines of equal ages
are shown in the figure, ranging from $10^7$ years (steepest slope)
via $5 \times 10^7$ years to $10^8$ years.

There is a marked difference between the two spiral galaxies and the
dwarfs. The \ion{H}{1} shells in M~31 and M~33 are smaller and the
range of expansion velocities is larger. In fact, when looking at the
objects in order of decreasing mass (M~31 -- M~33 -- IC~2574 -- Ho~II)
there seems to be a gradually shifting pattern. Whereas in the larger
spirals there is an absence of older holes (older than a few $\times
10^7$ years, holes in IC~2574 range between $5 \times 10^7$ and $10^8$
years and holes in Ho~II cluster near $10^8$ years

This suggests an alternative explanation for why we don't see younger
holes in IC~2574 and Ho~II. The SF histories of these dwarf galaxies
could be quite different from the spirals. In the latter, the global
SF rate is, to first order, constant with time.  Therefore, there is
always a population of young holes. In contrast, in gas--rich dwarf
irregulars, SF occurs in bursts which transforms them into an
\ion{H}{2} or blue compact dwarf (BCD) galaxy.  It might be that
IC~2574 and Ho~II have undergone a recent burst, perhaps triggered by
a close passage with another member of the M~81 group, and are now in
their post--starburst phase and that this is why we don't see many
young holes. Our data alone don't allow us to choose between the two
possibilities.

\placetable{diff}

\placefigure{massc}

\placefigure{energyc}

\placefigure{diamc}

\placefigure{agec}

\placefigure{velc}

\placefigure{v_d_comp}

Oey \& Clarke \markcite{OEY97}(1997), using the standard, adiabatic
shell evolution predict the size distribution for the populations of
OB superbubbles. In their paper they compare their predictions with
observations of the SMC and the three same galaxies as we compared our
IC~2574 data with, namely M31, M33, and Ho~II. Without going into
detail, their prediction fits the observations for the SMC
surprisingly well. This despite the fact that they ignore the
selection effect (missing short spacings) which renders the SMC maps
insensitive to large structures which should dominate the appearance
of dwarf galaxies. Also, they emphasize stalled shells, i.e., objects
which have reached their largest diameter. This is in contradiction
with the fact that the SMC shells, especially the smaller ones, are
all found to be rapidly expanding. Their approach works less well for
the other objects. They ascribe this to the data being incomplete, not
only when considering the smallest detectable shells, but overall, due
to the reduced sensitivity as compared to the SMC data. In that
respect, our IC~2574 are not better than the maps on M31, M33 or
Ho~II. Hence, a thorough test of Oey \& Clarke's predictions will have
to await higher resolution and higher sensitivity data on the nearest
(dwarf) galaxies.

\subsection{The intrinsic shape of the shells}
\label{shape}

One of the results from our analysis is that over 90\% of the holes are
well approximated by a circular shape (the few exceptions are
discussed in Sec.~\ref{single}).  This is a major difference with
other galaxies studied thus far in similar detail (i.e., Ho~II, M~31,
M~33, M~101, NGC~6946) which tend to have much more elliptical holes.
One attempt to understand this is in terms of projection effects
(different inclinations for the galaxies observed thus far). If the
shells are intrinsically triaxial, their projected shape on the sky
will invariably be an ellipse with an ellipticity which varies as a
function of the intrinsic axial ratios, the inclination of the galaxy,
{\em and} the position angle of the major axis of the hole within the
plane of the galaxy. This should lead to a wide range of
ellipticities. The fact that we don't see this in IC~2574 can only
mean that the shells in this galaxy are {\em intrinsically}
spherical. 

This can be understood by considering the shapes of the
rotation curves of the galaxies. IC~2574 is the only object which
exhibits a rising rotation curve throughout, as is typical for dwarf
galaxies. As the rotation curve does not turn over, leading to a flat
part dominated by differential rotation, there is no shear to distort
and ultimately destroy any large scale structures such as \ion{H}{1}
shells. For a more complete discussion on this topic, see Silich \ea
\markcite{SIL96}(1996) and Palou\v{s} \ea \markcite{PAL90}(1990). Indeed,
PWBR92\markcite{PUC92} estimate that the amount of shear over 1 or 2 kpc
in radius, once the flat part of the rotation curve is reached in
Ho~II, is sufficient to stretch the holes to an axial ratio of 0.5 over
a timescale of $10^8$ years. 
IC~2574, however, is the only object studied to date which
shows solid body rotation almost throughout the disk
(MCR94\markcite{MAR94}) which ensures that holes stay circular until
long after their creation.

There is one fly in the ointment. Why should the shells in IC~2574 be
intrinsically spherical? The expansion rate perpendicular to the disk
should be different from that in the disk. In the direction
perpendicular to the disk, the volume density rapidly decreases,
favoring faster expansion in that direction. But on the other hand,
the gas which moves upwards starts to feel the gravitational pull from
the mass surface density below it, slowing it down. It thus seems as
if nature conspires to create, at least in this particular object,
holes which are near spherical in shape.

Based on the absence of elliptical holes, an interesting conclusion
can be drawn concerning magnetic fields. Usually, the influence of
magnetic fields is neglected in numerical studies of expanding bubbles
in the ISM. However, a few analytical approaches have been put
forward in recent years to investigate the effects of the evolution of
shell--like structures in the presence of magnetic fields (e.g.,
Tomisaka \markcite{TOM91}1991). The results of these investigations are
that if global magnetic fields are present, this would inhibit the
expansion of a supernova driven bubble in directions perpendicular to
the field lines. The expansion along the field lines would not be
affected, thus leading to elliptical, elongated structures. The fact
that we observe only spherical holes in IC~2574 shows that either
magnetic fields play a minor role or that the magnetic fields in
IC~2574 are really weak.

\section{Summary}
\label{summa}

High resolution \ion{H}{1} and H$\alpha$ imaging using the VLA and the
Calar Alto observatory of the dwarf galaxy IC~2574 have been presented
in this paper.  The main results and conclusions are:

1. IC~2574 is a gas--rich dwarf galaxy which shows a stunning amount
of detail in the form of \ion{H}{1} shells and holes in its
interstellar medium.  These features are similar to those found in the
Galaxy, the Magellanic Clouds, M~31, M~33, M~101 and NGC~6946, IC~10
and Ho~II.  From the global \ion{H}{1} profile we derive a total
\ion{H}{1} mass of $9.0\times 10^8\,\mbox{M}_{\odot}$.

2. Using two independent approaches, the thickness of the \ion{H}{1}
layer of IC~2574 has been derived to be $\approx 350$ pc. This implies
that IC~2574, as is the case for Ho~II, is a 'thick' dwarf galaxy
(compared to more massive spirals) which can be understood in terms of
its lower gravitational potential.

3. Large scale structures emptied of \ion{H}{1} (believed to be
expanding \ion{H}{1} shells) leave roughly circular footprints,
\ion{H}{1} holes. They are prominent in the high resolution VLA
channel maps of IC~2574. In total, 48 \ion{H}{1} holes have been
catalogued which are distributed across the entire galaxy. The
diameters of the holes range between 100 pc and 1200 pc. The lower
limit of the sizes is due to the limit of our spatial resolution. Most
of the holes are found to be expanding with typical expansion
velocities ranging between 8 and 12 \kms.  The indicative ages of the
\ion{H}{1} holes range between 10 and $60\times 10^6$ years. Typical
energies that are required to create the observed features are of the
order of 10--100$\times 10^{50}$~ergs assuming that they were created
by the combined effects of stellar winds and supernova explosions. The
\ion{H}{1} masses that were present at the position of the holes
before creation (and that might now be present on the rim of the
shell) are of the order of 10--100$\times 10^4$~M$_{\odot}$.

4. In addition to the well defined holes, some large scale ($> 1000$
pc), coherent features are visible in the channel maps. They might be
the remainder of an older shell population.

5. The radial expansion of the holes, the indicative ages and the
energy requirements for their formation suggest that they have been
created by the combined effects of stellar winds and multiple
supernova explosions. Despite their large dimensions, the energy
requirements are not extreme and infall of material such as clouds is
not needed to explain the observed features. However, alternative
mechanisms cannot be ruled out.

6. The shells can grow to these large dimensions because of several
conditions which are fulfilled in dwarf galaxies. The volume density
in the plane is low, which facilitates expansion. In the
direction perpendicular to the disk, the gravitational pull is smaller
than in a massive spiral. Also, because of the thick \ion{H}{1}
layer, shells are easily contained and unlikely to blow--out. Lastly,
solid body rotation and a lack of spiral density waves prevent holes
from being rapidly destroyed.

7. A comparison with other galaxies studied in similar detail so far
(M~31, M~33 and Ho~II) shows that the size distribution of \ion{H}{1}
holes found in a galaxy is related to its Hubble type in the following
way. The size of the largest \ion{H}{1} shells is inversely
proportional to the global gravitational potential (and hence mass
surface density). The shells in dwarf galaxies show lower expansion
velocities, most likely due to the fact that the shells are preserved
much longer due to the absence of shear (differential rotation).  The
energies needed to create these structures, though, are found to be
roughly the same for all types of galaxies. If we follow the
conventional view that star forming regions are responsible for the
structures found in \ion{H}{1}, the energy output of a typical
star forming region does not seem to be related to the host galaxy.

8. The fact that virtually all holes in IC~2574 are circular in shape
is a major difference with the other galaxies studied in comparable
detail and can be explained, again, by the absence of shear and spiral
density waves in IC~2574. This finding also indicates that magnetic
fields play only a minor role in shaping expanding holes or that the
magnetic fields in IC~2574 are weak.

9. Current star formation, as traced by H$\alpha$ emission, is
predominantly found along the rims of the larger \ion{H}{1}--holes,
indicating propagating star formation.  A detailed comparison of the
H$\alpha$ and the \ion{H}{1} properties yields that, on a global
scale, star formation is only present if the \ion{H}{1} surface
density reaches values larger then $10^{21}$~cm$^{-2}$ (on a linear
scale of $\approx 95$~pc).

In summary, our results on IC~2574 confirm the findings by PWBR92 and
show that Ho~II is not a special case. Our analysis, and the
comparison with other galaxies studied in similar detail suggest that
there is a trend of hole characteristics with Hubble type (or
mass). If the conclusion that the energy output of a SF region is
independent of the host galaxy holds then it follows that there is a
lower mass limit for the formation of dwarf galaxies. In other words,
if the mass of a galaxy falls below a certain threshold, the first
burst of star formation will tear the whole galaxy apart (see also,
e.g., Puche \& Westpfahl \markcite{PUC94}1994). Mac~Low \& Ferrara
\markcite{MAF98}(1998) predict that this lower mass limit will be near
$10^6$~M$_{\odot}$, which is well below the mass of the objects dealt
with in this paper.

%
%
%
%

\acknowledgments

The authors are indebted to Mordecai--Marc Mac~Low for providing them
with a model data cube of a turbulent ISM in advance of publication and
to Johan van Horebeek for valuable advise regarding some of the
statistical techniques employed.  FW acknowledges the 'Deutsche
Forschungsgemeinschaft (DFG)' for the award of a stipendium in the
framework of the Graduiertenkolleg on "The Magellanic Clouds and other
Dwarf Galaxies" and would like to thank Uli Klein for continuous
support and Neb Duric for fruitful discussions. FW appreciates the
help of the staff of the Array Operation Center of the NRAO during his
stay in Socorro, New Mexico and also wants to thank the staff of the
Calar Alto observatory in Spain for excellent working conditions. FW
is grateful to Jan Palou\v s for illuminating discussions regarding
the numerical simulations of holes and shells during his stays at the
Academy of Science in Prague, Czech Republic. EB acknowledges a grant
awarded by CONACyT (grant number 0460P--E) and is grateful for the
support by the Graduiertenkolleg without which this collaborative
effort would not have been possible. Last but not least we would like
to thank Klaas de Boer for carefully reading the manuscript and an
anonymous referee for useful suggestions.  This research has made use
of the NASA/IPAC Extragalactic Database (NED) which is operated by the
Jet Propulsion Laboratory, Caltech, under contract with the National
Aeronautics and Space Administration (NASA), NASA's Astrophysical Data
System Abstract Service (ADS), and NASA's SkyView.

%
%
%
%

\newpage

\figcaption[walter.fig1n.eps]{R--band CCD--image of IC~2574 obtained
with the Calar Alto 2.2--m telescope.
\label{ic_r}}

\figcaption[walter.fig2n.eps]{Continuum--subtracted H$\alpha$ image of
IC~2574 obtained with the Calar Alto 2.2--m telescope. The total
exposure time is 45 minutes. The brightest stars that were not fully
subtracted by removing the continuum were blanked.
\label{ic_ha}}

\figcaption[walter.fig3an.ps]{Mosaic of every second channel map of
IC~2574 in which line emission was detected. The velocity of each
channel in \kms\ is plotted in the right hand corner. Greyscale range
from 0 (white) to 6 mJy~beam$^{-1}$ (black). Note that the beamsize
($6.4''\times 5.9''$ as indicated in the box of the top left hand
panel) is smaller than the resolution of this plot. Coordinates are
with respect to B1950.0 equinox.
\label{ic_cm_1}}

\figcaption[walter.fig4n.ps]{Global \ion{H}{1} profile for IC~2574
(squares) obtained by integrating the channel maps over the area of
the source after subtraction of the continuum background. The data
have been flux corrected (see text) and are corrected for primary beam
attenuation. The Flux F is given in Jy. The channel spacing is 2.58
\kms. For comparison, the profiles published by Rots (1980) (full
line) and MCR94 (dotted line) are also plotted. The arrow indicates
the velocity range affected by contamination due to galactic emission.
\label{ic_glob}} 

\figcaption[walter.fig5n.ps]{\ion{H}{1}\ surface brightness map of
IC~2574. The greyscale is a linear representation of the \ion{H}{1}
surface brightness ($0-2406$ K km s$^{-1}$). The beamsize
($6.4''\times5.9''$) is indicated in the lower left. 
\label{ic_mom0}}

\figcaption[walter.fig6n.ps]{Isovelocity contours of IC~2574 of the
$30''$ convolved cube. The velocities are given in \kms.  The
greyscale image is a linear representation of the \ion{H}{1} surface
brightness map at full resolution. A circular 30$''$ beam is plotted
in the lower left hand corner.
\label{ic_mom1}}

\figcaption[walter.fig7n.ps]{Velocity dispersion (second moment) map of
IC~2574. The greyscale is a linear representation of the velocity
dispersion ranging between 4 \kms\ (white) and 13 \kms\ (black). The
beamsize ($6.4''\times5.9''$) is indicated in the lower left.
\label{ic_mom2}}

\figcaption[walter.fig8n.ps]{Typical examples of position velocity
diagrams. a) pV diagram of a Type 3 hole (hole number 21). b) rV
diagram of the same hole. c) pV diagram of a Type 1 hole (number 30)
and d) pV diagram of a Type 2 hole (number 33).
\label{pvfinal}}

\figcaption[walter.fig9n.ps]{Location of the \ion{H}{1} holes in
IC~2574 (with numbers). The greyscale map is a linear representation
of the \ion{H}{1} surface brightness map. For the largest holes (13,
23, and 26), see text.
\label{ic_holes}}

\figcaption[walter.fig10n.ps]{H$\alpha$ image of IC~2574 overlaid with
the the discovered holes in its neutral interstellar medium (with
numbers).  Regions A--C are enlarged in Figs.~\ref{har1}--\ref{har3},
respectively. Note that almost all H$\alpha$ emission in IC~2574 is
located on the rims of \ion{H}{1} shells (see text).
\label{haho}}

\figcaption[walter.fig11n.ps]{H$\alpha$ image of IC~2574 overlaid with
one contour of the \ion{H}{1} surface brightness map at $1.7\times
10^{21}$ atoms cm$^{-2}$. Note that regions where we find H$\alpha$
emission coincide with high \ion{H}{1} surface brightness.
\label{hahic}}

\figcaption[walter.fig12n.ps]{Enlargement of region A from
Fig.~\ref{haho}.  The circles indicate where we detected the
\ion{H}{1} holes (for identification of the \ion{H}{1} holes, please
refer to Fig.~\ref{haho}). The boxes show how we defined groups of
\ion{H}{2} regions within IC~2574 (with sequence numbers).
\label{har1}}

\figcaption[walter.fig13n.ps]{Enlargement of region B from
Fig.~\ref{haho}.  The circles indicate where we detected the
\ion{H}{1} holes (for identification of the \ion{H}{1} holes, please
refer to Fig.~\ref{haho}). The boxes show how we defined groups of
\ion{H}{2} regions within IC~2574 (with sequence numbers).
\label{har2}}

\figcaption[walter.fig14n.ps]{Enlargement of region C from
Fig.~\ref{haho}.  The circles indicate where we detected the
\ion{H}{1} holes (for identification of the \ion{H}{1} holes, please
refer to Fig.~\ref{haho}). The boxes show how we defined groups of
\ion{H}{2} regions within IC~2574 (with sequence numbers).
\label{har3}}

\figcaption[walter.fig15n.ps]{Comparison of the relative distribution,
in percentage, of the indicative masses of the \ion{H}{1} holes in
Ho~II, IC~2574, M~33 and M~31.
\label{massc}}

\figcaption[walter.fig16n.ps]{Comparison of the relative distribution,
in percentage, of the energies required to produce the \ion{H}{1}
holes in Ho~II, IC~2574, M~33 and M~31.
\label{energyc}}

\figcaption[walter.fig17n.ps]{Comparison of the relative distribution,
in percentage, of the diameters of the \ion{H}{1} holes in Ho~II,
IC~2574, M~33 and M~31.
\label{diamc}}  

\figcaption[walter.fig18n.ps]{Comparison of the relative distribution,
in percentage, of the ages of the \ion{H}{1} holes in Ho~II, IC~2574,
M~33 and M~31.
\label{agec}}

\figcaption[walter.fig19n.ps]{Comparison of the relative distribution,
in percentage, of the expansion velocities of the \ion{H}{1} holes in
Ho~II, IC~2574, M~33 and M~31.
\label{velc}}

\figcaption[walter.fig20n.ps]{The relation between expansion velocity
and diameter in each of the four galaxies. Crosses represent observed
values. The plotted curves are lines of constant energy (based on
Chevalier 1974) at levels of $10^{49}$, $10^{50}$, $10^{51}$ and
$10^{52} $ergs (from left to right). The straight lines correspond to
constant age and are drawn at $10^7$ years (steepest slope), $5 \times
10^7$ years and $10^8$ years.
\label{v_d_comp}}

%
%
%
%

\newpage

\begin{table*}
\caption{General information on IC~2574 \label{ic_info}}
\tablenum{1}
\begin{center}
\begin{tabular}{lc}
\tableline \tableline
Object & IC~2574 \nl
Other names & UGC~5666, DDO~81, VII Zw 330 \nl
Right ascension (1950.0)\tablenotemark{a} & 10$^{\rm h}$24$^{\rm m}$41\fs3 \nl
Declination (1950.0)\tablenotemark{a} & 68$^{\circ}$ 40$'$ 18$''$\nl
Adopted distance\tablenotemark{b} & 3.2 Mpc \nl
Inclination\tablenotemark{c} & 77$\pm$3$^\circ$ \nl 
Scale (1$''$ corresponds to) & 15.5 pc \nl
Apparent magnitude (B)\tablenotemark{d} & 10.8 mag\nl
Corrected distance modulus & m$_{\rm B}$-M$_{\rm B}$=27.6 \nl
Corrected absolute magnitude (B) & -16.8 mag \nl
Blue luminosity L$_{\rm B}$ & 7.8$\times10^8$ L$_{\rm B\odot}$ \nl
HI mass M$_{\rm HI}$\tablenotemark{e} & 9.0$\times10^8$ M$_{\odot}$ \nl
HI mass to blue light ratio M$_{\rm HI}$/L$_{\rm B}$ & 1.15 M$_{\odot}$/L$_{\rm B\odot}$\nl
\tableline
\tableline
\end{tabular}
\tablenotetext{a}{Dressel \& Condon \markcite{DRE76}1976}
\tablenotetext{b}{see discussion in Puche \ea \markcite{PUC92}1992}
\tablenotetext{c}{Martimbeau \ea \markcite{MAR94}1994}
\tablenotetext{d}{de Vaucouleurs {\em et al.} \markcite{VAU91}1991 (RC3)}
\tablenotetext{e}{this study}
\end{center}
\end{table*}

\newpage

\begin{table*}
\tablenum{2}
\caption{Setup of the VLA during the observations}
\label{VLAsetup}
\begin{center}
\small
\begin{tabular}{lccc}
\tableline \tableline
VLA configuration & B & C & D \nl
baselines & $1-54\,\mbox{k}\lambda$ & $0.34 - 16\,\mbox{k}\lambda$ & $0.166 - 4.
9\,\mbox{k}\lambda$\nl
& $ (0.21 - 11.4 \,\mbox{km})$ &  $ (0.073 - 3.4 \,\mbox{km})$ & $ (0.035 - 1.03
\,\mbox{km})$\nl
date of observation & 1992 Jan.\ 18 & 1992 Apr.\ 20 & 1991 Mar.\ 03  \nl
                    & 1992 Jan.\ 19 &               & 1992 Sep.\ 03  \nl 
total time on source & 647 min & 133 min & 245 min \nl
\tableline
total bandwidth &  & 1.56 MHz &  \nl
No. of channels &  & 128 &  \nl
velocity resolution &  & 2.58 \kms &  \nl
central velocity &  & 38 \kms &  \nl
angular resolution\tablenotemark{a} & & $6.4 \times 5.9$ arcsec\nl
linear resolution\tablenotemark{a} &  & $100 \times 91$ pc\nl
rms noise\tablenotemark{a} & & 0.7 mJy\,$\times$\,beam$^{-1}$ \nl
        & & 11 Kelvin\nl
\tableline \tableline
\end{tabular}
\tablenotetext{a}{based on {\sc robust} data cube; see text}
\end{center}

\end{table*}

\newpage

\begin{table*}
\tablenum{3}
\caption{Observed properties of the \ion{H}{1} holes}
\label{holecat}
\begin{center}
\scriptsize
\begin{tabular}{rccccccccc}
\tableline \tableline
$\#$ & $\alpha$ & $\delta$ & $v_{\rm hel}$ & d & $v_{\rm exp}$ & $n_{\rm HI}$ & type \nl
     & (1950.0) & (1950.0) & (\kms) & (pc) & (\kms) & (cm$^{-3}$) & \nl
\tableline
1 & 10 23 34.5 & 68 36 26  & -11   & 310   & 8     & 0.13  & 3     \nl
2 & 10 23 38.0 & 68 36 43  & -1  & 310   & 5     & 0.13  & 3 \nl
3 & 10 23 48.0 & 68 36 58  & 5   & 976   & 17    & 0.15  & 3  \nl
4 & 10 23 52.5 & 68 40 14  & 28  & 1240$\times$1116,42$^\circ$   & 14    & 0.05  & 2 \nl
5 & 10 23 59.0 & 68 34 07  & -1  & 775   & 9     & 0.1   & 3 \nl
6 & 10 24 00.5 & 68 42 08  & 53  & 325   & 8     & 0.04  & 3  \nl
7 & 10 24 02.5 & 68 36 50  & 2   & 1162  & --    & 0.16  & 1 \nl
8 & 10 24 04.6 & 68 39 57  & 33    & 372         & 15    & 0.08  & 3 \nl
9 & 10 24 05.5 & 68 37 59  & 17  & 418   & 7     & 0.16  & 3 \nl
10 & 10 24 06.5 & 68 36 40  & 7  & 418   & --    & 0.15  & 1 \nl
11 & 10 24 12.0 & 68 42 33 & 43  & 697   & 8     & 0.04  & 3 \nl
12 & 10 24 15.0 & 68 37 12 & 15  & 325   & --    & 0.16  & 1 \nl
13 & 10 24 15.0 & 68 39 13 & 25  & 1922  & --    & 0.16  & 1 \nl
14 & 10 24 17.0 & 68 38 11 & 20  & 325   & --    & 0.17  & 1 \nl
15 & 10 24 23.5 & 68 37 20 & 25  & 697   & 6     & 0.16  & 3 \nl
16 & 10 24 26.5 & 68 37 57 & 25  & 325   & 8     & 0.16  & 3 \nl
17 & 10 24 26.5 & 68 43 33 & 53  & 465   & 13    & 0.08  & 3 \nl
18 & 10 24 28.0 & 68 38 43 & 28  & 434   & 11    & 0.17  & 3 \nl
19 & 10 24 31.5 & 68 42 58 & 36  & 232   & 8     & 0.07  & 3 \nl
20 & 10 24 42.0 & 68 40 18 & 33  & 837   & --    & 0.16  & 1 \nl
21 & 10 24 35.0 & 68 39 48 & 38  & 604   & 18    & 0.17  & 3 \nl
22 & 10 24 40.5 & 68 39 35 & 38  & 976$\times$558,78$^\circ$     & 13    & 0.17  & 3 \nl
23 & 10 24 41.0 & 68 42 35 & 46  & 1550  & --    & 0.11  & 1 \nl
24 & 10 24 42.0 & 68 40 18 & 46  & 325   & --    & 0.17  & 1 \nl
25 & 10 24 44.0 & 68 35 57 & 25  & 775   & 13    & 0.05  & 3 \nl
26 & 10 24 48.5 & 68 37 08 & 33  & 1240  & --    & 0.07  & 1 \nl
27 & 10 24 49.0 & 68 41 09 & 48  & 744   & --    & 0.17  & 1 \nl
28 & 10 24 49.5 & 68 39 43 & 53  & 279   & 6     & 0.17  & 3 \nl
29 & 10 24 52.5 & 68 43 22 & 56  & 356   & 10    & 0.09  & 3 \nl
30 & 10 24 56.0 & 68 39 59 & 53  & 1085  & --    & 0.17  & 1 \nl
31 & 10 24 57.0 & 68 43 19 & 51  & 186   & --    & 0.12  & 1 \nl
32 & 10 25 00.5 & 68 41 43 & 61  & 604   & 24    & 0.17  & 3 \nl
33 & 10 25 01.5 & 68 42 49 & 66  & 744   & 23    & 0.15  & 2 \nl
34 & 10 25 02.5 & 68 42 26 & 48  & 232   & 9     & 0.16  & 3 \nl
35 & 10 25 03.0 & 68 43 33 & 64  & 1023$\times$496,74$^\circ$    & 25    & 0.13  & 2 \nl
36 & 10 25 05.0 & 68 44 55 & 82  & 418   & 10    & 0.07  & 3 \nl
37 & 10 25 05.5 & 68 44 19 & 77  & 511   & --    & 0.07  & 1 \nl
38 & 10 25 06.5 & 68 40 28 & 69  & 310   & --    & 0.15  & 1 \nl
39 & 10 25 08.0 & 68 41 39 & 64  & 697   & --    & 0.17  & 1 \nl
40 & 10 25 11.5 & 68 44 01 & 77  & 232   & 10    & 0.12  & 3 \nl
41 & 10 25 13.0 & 68 38 45 & 59  & 837   & --    & 0.06  & 1 \nl
42 & 10 25 14.5 & 68 43 57 & 77  & 279   & 13    & 0.13  & 2 \nl
43 & 10 25 15.0 & 68 37 57 & 56  & 697$\times$418,80$^\circ$& --         & 0.05  & 2 \nl
44 & 10 25 16.5 & 68 37 53 & 56  & 186   & 8     & 0.05  & 3 \nl
45 & 10 25 18.5 & 68 41 13 & 87  & 418   & 7     & 0.16  & 3 \nl
46 & 10 25 21.8 & 68 41 57 & 87  & 155   & 4     & 0.16  & 3 \nl
47 & 10 25 28.5 & 68 41 13 & 71  & 155   & 8     & 0.05  & 3 \nl
48 & 10 25 34.0 & 68 42 54 & 92  & 403$\times$186,80$^\circ$     & 8     & 0.15  & 2 \nl
\tableline
\tableline
\end{tabular}
\end{center}
\end{table*}

\newpage

\begin{table*}
\tablenum{4}
\caption{Derived properties of the \ion{H}{1} holes}
\label{propcat}
\begin{center}
\scriptsize
\begin{tabular}{rccccc}
\tableline \tableline
$\#$ & d & $v_{exp}$ & age $t$ & mass $M_{\rm HI}$ & energy $E_{\rm E}$ \nl
     & (pc) & (\kms) & ($10^6$ yr) & ($10^4$ M$_\odot$) & ($10^{50}$ ergs) \nl
\tableline
1 &     310 & 8          & 19.6  & 4.9   & 6.8 \nl
2 &     310 &   5        & 31.3  & 4.9   & 3.5\nl
3 &     976.5  & 17      & 29.0  & 178.4 & 817.9\nl
4 &     1178 &  14       & 42.5  & 104.4 & 326.9\nl
5 &     775 &   9        & 43.5  & 59.5  & 103.7\nl
6 &     325.5 &  8       & 20.6  & 1.8   & 2.1\nl
7 &     1162.5 & --      & --    & 321.1 & --\nl
8 &     372    & 15      & 12.2  & 5.2     & 16.7\nl
9 &     418.5 & 7        & 30.2  & 15.0  & 18.1\nl
10 &    418.5 & --       & --    & 14.0  & --\nl
11 &    697.5 & 8        & 44.1  & 17.3  & 22.7\nl
12 &    325.5 & --       & --    & 7.0 & --\nl
13 &    1922 &  --       & --    & 1451.2& --\nl
14 &    325.5 & --       & --    & 7.5   & --\nl
15 &    697.5 & 6        & 58.7  & 69.4  & 71.6\nl
16 &    325.5 & 8        & 20.6  & 7.0   & 9.9\nl
17 &    465 &   13       & 18.1  & 10.3  & 27.4\nl
18 &    434 &   11       & 19.9  & 17.8  & 40.7\nl
19 &    232.5 & 8        & 14.7  & 1.1   & 1.4\nl
20 &    837 &   --       & --    & 119.9 & --\nl
21 &    604.5 & 18       & 17.0  & 48.0  & 228.3\nl
22 &    744 &   13       & 28.9  & 89.4  & 276.7\nl
23 &    1550 &  --       & --    & 523.3 & --\nl
24 &    325.5 & --       & --    & 7.5   & --\nl
25 &    775 &   13       & 30.1  & 29.7  & 79.8\nl
26 &    1240 &  --       & --    & 170.5 & --\nl
27 &    744 &   --       & --    & 89.4  & --\nl
28 &    279 &   6        & 23.5  & 4.7   & 4.4\nl
29 &    356.5 & 10       & 18.0  & 5.2   & 9.5\nl
30 &    1085 &  --       & --    & 277.4 & --\nl
31 &    186 &   --       & --    & 1.0   & --\nl
32 &    604.5 & 24       & 12.7  & 48.0  & 341.5\nl
33 &    744 &   23       & 16.3  & 78.9  & 534.6\nl
34 &    232.5 & 9        & 13.1  & 2.6   & 4.1\nl
35 &    713 &   25       & 14.3  & 60.2  & 420\nl
36 &    418.5 &  10      & 21.1  & 6.6   & 11.8\nl
37 &    511.5 & --       & --    & 12.0  & --\nl
38 &    310 &   --       & --    & 5.7   & --\nl
39 &    697.5 & --       & --    & 73.7  & --\nl
40 &    232.5 & 10       & 11.7  & 1.9   & 3.4\nl
41 &    837 &   --       & --    & 44.9  & --\nl
42 &    279 &   13       & 10.8  & 3.6   & 9.6\nl
43 &    542.5 & --       & --    & 10.2  & --\nl
44 &    186 &   8        & 11.7  & 0.4   & 0.5\nl
45 &    418.5 & 7        & 30.2  & 15.0  & 18.1\nl
46 &    155 &   4        & 19.6  & 0.8   & 0.4\nl
47 &    155 &   8        & 9.8   & 0.2   & 0.3\nl
48 &    279 &   8        & 17.6  & 4.2   & 5.7\nl
\tableline
\tableline
\end{tabular}
\end{center}
\end{table*}

\newpage

\begin{table*}
\tablenum{5}
\caption{Cross identification of our grouping of \ion{H}{2} regions
with the  ones defined by MH94}
\label{cross}
\begin{center}
\scriptsize
\begin{tabular}{rccl}
\tableline \tableline
Number & $\alpha_{1950.0}$ & $\delta_{1950.0}$  & cross identifications \nl
       &                   &                   & Miller \& Hodge 1994 (MH94) \nl
\tableline
1$'$  & 10 25 20.2 & 68 41 40  & 246,251 \nl
2$'$  & 10 25 10.1 & 68 40 50  & 181,190,193,197,203,206,\nl
    &            &           & 207,218,221,226,227,230 \nl
3$'$  & 10 25 10.2 & 68 40 32  & 208 \nl
4$'$  & 10 25 10.4 & 68 40 19  & 195,213,214,227 \nl
5$'$  & 10 25 05.7 & 68 41 15  & 184 \nl
6$'$  & 10 25 04.5 & 68 39 39  & 182,174 \nl
7$'$  & 10 24 59.5 & 68 41 51  & 136,143,155,158,165 \nl
8$'$  & 10 24 54.1 & 68 40 53  & 125,126,131 \nl
9$'$  & 10 24 48.9 & 68 40 31  & 117 \nl
10$'$ & 10 24 49.9 & 68 40 18  & 120 \nl
11$'$ & 10 24 49.5 & 68 39 43  & 115,121 \nl
12$'$ & 10 24 48.7 & 68 39 27  & 113,116 \nl
13$'$ & 10 24 41.3 & 68 38 58  & 99,100,106 \nl
14$'$ & 10 24 44.5 & 68 40 07  & 111 \nl
15$'$ & 10 24 42.0 & 68 40 37  & 105 \nl
16$'$ & 10 24 40.1 & 68 40 05  & 94,96,102,104 \nl
17$'$ & 10 24 33.1 & 68 39 33  & 76,80,83,84 \nl
18$'$ & 10 24 20.4 & 68 37 55  & 57,58,60-63,65,67 \nl
19$'$ & 10 24 13.9 & 68 37 44  & 50 \nl
20$'$ & 10 24 09.3 & 68 37 24  & 46 \nl
21$'$ & 10 23 57.0 & 68 38 37  & 17,19,20 \nl
22$'$ & 10 24 02.7 & 68 39 12  & 30,33 \nl
23$'$ & 10 24 02.5 & 68 39 23  & 31 \nl
24$'$ & 10 24 00.1 & 68 39 17  & 25 \nl
25$'$ & 10 24 04.5 & 68 39 58  & 24,28,32,34,36,38-41,43-46 \nl
26$'$ & 10 23 59.7 & 68 39 58  & 24 \nl
27$'$ & 10 24 49.1 & 68 43 30  & 118 \nl
28$'$ & 10 24 58.4 & 68 42 29  & 144 \nl
29$'$ & 10 25 05.7 & 68 42 35  & 186 \nl
30$'$ & 10 25 03.9 & 68 43 04  & 178 \nl
31$'$ & 10 24 59.4 & 68 43 23  & 148 \nl
32$'$ & 10 24 56.1 & 68 43 24  & 133,134,137,138 \nl 
33$'$ & 10 24 58.0 & 68 43 52  & 140 \nl
34$'$ & 10 25 03.0 & 68 43 51  & 167,169,179,180 \nl
35$'$ & 10 25 07.2 & 68 43 26  & 192,196 \nl
36$'$ & 10 25 08.4 & 68 43 49  & 194,198,202 \nl
37$'$ & 10 25 12.0 & 68 44 06  & 210,215,224,225 \nl
38$'$ & 10 25 14.8 & 68 43 51  & 235 \nl
39$'$ & 10 25 17.9 & 68 43 51  & 237,238,240,242,244,245 \nl
40$'$ & 10 25 15.7 & 68 43 20  & 233,237,240 \nl
\tableline
\tableline
\end{tabular}
\end{center}
\end{table*}

\newpage

\begin{table*}
\tablenum{6}
\caption{Derived properties of grouped \ion{H}{2} regions}
\label{Ha-regions}
\begin{center}
\scriptsize
\begin{tabular}{rcccccccl}
\tableline \tableline
Number & $F$(H$\alpha$) & $\rho_{\rm HII}$ & r$_{\rm HI,eff}$ & M$_{\rm HI}$ & $\Delta v$ & $\rho_{\rm HI}$ & $\sigma_{\rm HI}$ & remarks \nl   
 & [$10^{-15}\!\!$ erg$\!$ cm$^{-2}$s$^{-1}$] & [cm$^{-3}$] & [pc] & [10$^5$ M$_{\odot}$] & km s$^{-1}$ & [cm$^{-3}$] & [10$^{20}$ cm$^{-2}$] &  \nl
\tableline
1$'$  & 30.8   & 0.7  & 201 & 14  & 16.3 & 1.68 & 13.9 & HI--clump\nl
2$'$  & 434.2  & 1.8  & 225 & 24  & 26.1 & 2.05 & 18.9 & HI--clump\nl
3$'$  & 32.9   & 0.85 & 117 & 5.8 & 26.3 & 3.52 & 16.9 & new tiny hole (D$<$60pc), rim hole 38\nl
4$'$  & 41.9   & 0.8  & 148 & 10  & 19.7 & 3.00 & 18.3 & diffuse HI\nl
5$'$  & 13.7   & 0.5  & 152 & 7.4 & 22.4 & 2.05 & 12.8 & on rim of hole 39, diffuse HI\nl
6$'$  & 34.5   & 0.8  & 245 & 21  & 18.3 & 1.39 & 14.0 & HI--clump\nl
7$'$  & 152.4  & 0.6  & 270 & 17  & 22.4 & 0.84 & 9.3  & H$\alpha$ fills hole 32!\nl
8$'$  & 55.2   & 0.7  & 196 & 14  & 23.7 & 1.81 & 14.6 & on rim of hole 27\nl
9$'$  & 5.2    & 0.8  & 111 & 4.8 & 26.3 & 3.41 & 15.6 & --\nl
10$'$ & 5.2    & 0.6  & 107 & 4.7 & 21.7 & 3.73 & 16.4 & on rim of hole 30\nl
11$'$ & 40.8   & 0.6  & 100 & 3.9 & 28.9 & 5.23 & 15.6 & H$\alpha$--em. on rim of hole 28\nl
12$'$ & 9.9    & 0.65 & 103 & 4.4 & 25.0 & 3.92 & 16.6 & new tiny hole (D$<$60pc)\nl
13$'$ & 44.7   & 0.6  & 295 & 28  & 23.7 & 1.06 & 12.9 & on rim of hole 22, diffuse HI\nl
14$'$ & 8.9    & 0.9  & 88  & 2.7 & 26.3 & 3.85 & 13.9 & on rim of hole 24, diffuse HI\nl
15$'$ & 9.9    & 0.6  & 155 & 7.9 & 20.0 & 2.06 & 13.1 & on rim of hole 24. diffuse HI\nl
16$'$ & 102.2  & 0.9  & 163 & 10  & 34.2 & 2.25 & 15.0 & on rim of holes 21,22,24, high $\Delta$v!\nl
17$'$ & 86.8   & 1.4  & 142 & 8.5 & 25.0 & 2.89 & 16.9 & on rim of holes 21,22\nl
18$'$ & 83.8   & 0.8  & 303 & 41  & 25.0 & 1.43 & 18.7 & a lot of HI in region\nl
19$'$ & 139.0  & 1.1  & 130 & 8.1 & 20.4 & 3.58 & 19.1 & a lot of HI in region\nl
20$'$ & 13.6   & 0.7  & 111 & 4.3 & 22.4 & 3.00 & 14.0 & a lot of HI in region\nl
21$'$ & 75.4   & 0.85 & 233 & 17  & 18.4 & 1.30 & 12.5 & new tiny hole (D$<$60pc)\nl
22$'$ & 27.5   & 1.0  & 77  & 2.4 & 19.6 & 5.11 & 16.2 & HI--clump between H$\alpha$--em.\nl 
23$'$ & 13.8   & 0.8  & 95  & 3.2 & 19.7 & 2.37 & 14.2 & diffuse HI\nl
24$'$ & 9.6    & 0.7  & 93  & 2.8 & 21.0 & 2.43 & 12.9 & diffuse HI\nl
25$'$ & 213.2  & 0.8  & 270 & 23  & 31.6 & 1.14 & 12.6 & H$\alpha$ same as rim of hole 8!\nl
26$'$ & 16.2   & 0.8  & 73  & 2.4 & 31.3 & 6.00 & 18.0 & on rim of holes 8,13, high $\Delta$v \nl
27$'$ & 35.9   & 0.9  & 124 & 6.2 & 25.0 & 3.16 & 16.1 & on rim of holes 23,29\nl
28$'$ & 7.8    & 0.6  & 89  & 2.7 & 21.0 & 3.73 & 13.6 & new tiny hole (D$<$60pc), rim hole 33\nl
29$'$ & 109.0  & 0.6  & 93  & 4.1 & 26.3 & 4.96 & 18.9 & diffuse HI, on rim of hole 33\nl
30$'$ & 55.0   & 0.8  & 136 & 7.1 & 23.7 & 2.75 & 15.3 & new tiny hole (D$<$60pc)\nl
31$'$ & 120.9  & 1.4  & 45  & 3.3 & 34.2 & 3.52 & 6.5  & hardly any HI, high $\Delta$v\nl
32$'$ & 204.3  & 1.5  & 175 & 14  & 33.0 & 2.54 & 18.3 & on rim of holes 31,35, high $\Delta$v\nl
33$'$ & 78.8   & 0.85 & 93  & 2.5 & 32.0 & 3.02 & 11.5 & little HI, high $\Delta$v\nl
34$'$ & 1004.7 & 2.0  & 199 & 8.9 & 25.0 & 1.03 & 9.0  & little HI, on rim of holes 35,37\nl
35$'$ & 444.9  & 2.0  & 154 & 11  & 23.5 & 2.93 & 18.5 & a lot HI, on rim of hole 35 \nl
36$'$ & 487.2  & 1.9  & 183 & 17  & 28.7 & 2.70 & 20.3 & a lot HI, rim hole 35, high $\Delta$v\nl
37$'$ & 140.4  & 1.3  & 172 & 13  & 35.3 & 2.48 & 17.5 & little HI, rim hole 40, high $\Delta$v,\nl
38$'$ & 13.3   & 0.8  & 98  & 5.0 & 23.7 & 5.17 & 20.8 & on rim of hole 42\nl
39$'$ & 144.8  & 1.2  & 207 & 28  & 22.4 & 3.07 & 26.1 & a lot of HI in region\nl
40$'$ & 15.4   & 0.85 & 182 & 15  & 25.6 & 2.42 & 18.1 & a lot of HI in region\nl
\tableline
\tableline
\end{tabular}
\end{center}
\end{table*}

\newpage

\begin{table*}
\tablenum{7}
\caption{Some properties of high mass stars (based on Devereux \ea 1997)} 
\label{sprop}
\begin{center}
\small
\begin{tabular}{cccccc}
\tableline \tableline
Type & Mass (M$_\odot$) & L$_{\rm bol}$ (L$_\odot$) & N$_{\rm Lyc}$ & L(H$\alpha$) & L$_{\rm bol}$/L(H$\alpha$) \nl
\tableline
O5 & 40 & $6.8\times 10^5$ & $4\times 10^{49}$   & $1.4\times 10^4$ & 49 \nl  
O7 & 25 & $1.0\times 10^5$ & $4\times 10^{48}$   & $1.4\times 10^3$ & 71 \nl
O9 & 20 & $4.6\times 10^4$ & $1.3\times 10^{48}$ & $4.6\times 10^2$ & 100 \nl
B0 & 16 & $2.5\times 10^4$ & $2.5\times 10^{47}$ & 89               & 282 \nl
B1 & 10 & $5.2\times 10^3$ & $1.9\times 10^{45}$ & 0.67 & $7.8\times 10^3$ \nl
B2 & 8  & $2.9\times 10^3$ & $4.5\times 10^{44}$ & 0.16 & $1.8\times 10^4$ \nl
B5 & 6  & $7.9\times 10^2$ & $4.8\times 10^{43}$ & 0.02 & $4.0\times 10^4$  \nl
\tableline \tableline
\end{tabular}
\end{center}
\end{table*}

\newpage

\begin{table*}
\tablenum{8}
\caption{Summary of \ion{H}{1} hole statistics in four nearby galaxies}
\label{comp}
\begin{center}
\small
\begin{tabular}{lrrrr}
\tableline \tableline
Property & M~31\tablenotemark{a} & M~33\tablenotemark{b} &
IC~2574\tablenotemark{c} & Ho~II\tablenotemark{d} \nl
\tableline
Linear resolution (pc) & 100 & 55 & 95  & 65 \nl  
Velocity resolution (km s$^{-1}$) & 8.2 & 8.2 & 2.6 & 2.6 \nl
Average surface density  ($10^{20}$ cm$^{-2}$) & 5 & 9 & 4 &10 \nl
Average volume density (cm$^{-3}$) & 0.6 & 0.45 & 0.15 & 0.2 \nl
Average velocity dispersion (km s$^{-1}$) & 8 & 8 & 7 &7 \nl
Derived scaleheight (pc) & 120 & 100 & 350 & 625 \nl
Number of holes & 141 & 148 & 48 & 51 \nl
Sensitivity per channel ($10^{20}$ cm$^{-2}$) & 0.3 & 1.0 & 0.5 &2.4 \nl
\tableline 
\tableline
\end{tabular}
\tablenotetext{a}{Brinks \& Bajaja \markcite{BRI86}1986}
\tablenotetext{b}{Deul \& den Hartog \markcite{DEU90}1990}
\tablenotetext{c}{This paper}
\tablenotetext{d}{Puche \ea  \markcite{PUC92}1992}
\end{center}

\end{table*}

\end{document}